\documentclass{aastex6}
\usepackage[hyperref]{} % For arxiv
\usepackage{epsfig}
\usepackage{psfig}
\usepackage{color}
\newcommand{\bsym}[1]{\mbox{\boldmath${#1}$}}

\newcommand{\grad}{\ensuremath{\nabla}}

\newcommand{\vect}[1]{\ensuremath{\bsym{#1}}}

\newcommand{\divgh}{\ensuremath{\nabla_h\cdot\,}}
\newcommand{\curlh}{\ensuremath{\nabla_h\times\,}}

\include{defs}
\newcommand{\be}{\begin{equation}}
\newcommand{\ee}{\end{equation}}
\newcommand{\bea}{\begin{eqnarray}}
\newcommand{\eea}{\end{eqnarray}}
\newcommand{\beax}{\begin{eqnarray*}}
\newcommand{\eeax}{\end{eqnarray*}}
\newcommand{\ba}{\begin{array}}
\newcommand{\ea}{\end{array}}
\newcommand{\bed}{\begin{description}}
\newcommand{\ed}{\end{description}}
\newcommand{\blc}{\begin{list}{$\circ$}{}}
\newcommand{\blb}{\begin{list}{$\bullet$}{}}
\newcommand{\el}{\end{list}}
\newcommand{\ben}{\begin{enumerate}}
\newcommand{\een}{\end{enumerate}}

\newcommand{\nn}{\nonumber}

\newcommand{\hatz}{\,\hat{\bsym{z}}}
\newcommand{\hatn}{\,\hat{\bsym{n}}}

\renewcommand{\aap}{    {\it Astron. Astrophys.}}

\renewcommand{\apj}{    {\it Astrophys. J.}}
\renewcommand{\apjl}{   {\it Astrophys. J. Lett.}}
\renewcommand{\apss}{   {\it Astrophys. Space Sci.}}

\renewcommand{\solphys}{{\it Solar Phys.}}

\renewcommand{\arcsec}{\hbox{$^{\prime\prime}$}}

\begin{document}
%\begin{article}
\tracingmacros=2
%\begin{opening}
\title{%
Deriving Potential Coronal Magnetic Fields from Vector Magnetograms
}
%\author[addressref={aff1,aff2},corref,email={welsch@ssl.berkeley.edu}]{\inits{B.}\fnm{Brian T.}~\lnm{Welsch}}%\sep
\author{Brian T.~Welsch\altaffilmark{1}}
\affil{Natural \& Applied Sciences, University of Wisconsin - Green
  Bay, 2420 Nicolet Dr., Green Bay, WI 54311}
\email{welschb@uwgb.edu}
\and
\author{George H.~Fisher}
\affil{Space Sciences Laboratory, 7 Gauss Way, University of California,
  Berkeley, CA 94720-7450}

\altaffiltext{1}{Space Sciences Laboratory, 7 Gauss Way, University of California,
  Berkeley, CA 94720-7450}

\shortauthors{Welsch \& Fisher}
\shorttitle{Potential Fields from Vector Magnetograms}

\begin{abstract}
The minimum-energy configuration for the magnetic field above the
solar photosphere is curl-free (hence, by Amp\`ere's law, also
current-free), so can be represented as the gradient of a scalar
potential. Since magnetic fields are divergence free, this scalar
potential obeys Laplace's equation, given an appropriate boundary
condition (BC). With measurements of the full magnetic vector at the
photosphere, it is possible to employ either Neumann or Dirichlet BCs
there. Historically, the Neumann BC was used with available
line-of-sight magnetic field measurements, which approximate the
radial field needed for the Neumann BC. Since each BC fully determines
the 3D vector magnetic field, either choice will, in general, be
inconsistent with some aspect of the observed field on the boundary,
due to the presence of both currents and noise in the observed
field. We present a method to combine solutions from both Dirichlet
and Neumann BCs to determine a hybrid, ``least-squares'' potential
field, which minimizes the integrated square of the residual between
the potential and actual fields. We also explore weighting the
residuals in the fit by spatially uniform measurement
uncertainties. This has advantages in both not overfitting the radial
field used for the Neumann BC, and maximizing consistency with the
observations. We demonstrate our methods with SDO/HMI vector magnetic
field observations of AR 11158, and find that residual discrepancies
between the observed and potential fields are significant, and are
consistent with nonzero horizontal photospheric currents. We also
analyze potential fields for two other active regions observed with
two different vector magnetographs, and find that hybrid potential
fields have significantly less energy than the Neumann fields in every
case --- by more than 10$^{32}$ erg in some cases. This has major
implications for estimates of free magnetic energy in coronal field
models, {\it e.g.}, non-linear force-free field extrapolations.
\end{abstract}
%\end{opening}
\tracingmacros=0

\section{Introduction}
\label{sec:intro}
Solar flares and coronal mass ejections (CMEs) are thought to be driven by the release of magnetic energy stored in electric currents in the solar corona ({\it e.g.}, \citealp{Forbes2000}). Consequently, the origins and structure of coronal electric currents are a subject of intense research. Currently, however, measurements of the coronal magnetic field are rare and subject to substantial uncertainties ({\it e.g.}, \citealp{Lin2004, Tomczyk2008}): the vector field as a function of three spatial coordinates cannot be directly measured. In contrast, vector magnetograms --- maps of the magnetic field vector over part of the solar photosphere, typically of active regions --- have been made for decades. These reveal electric currents normal to the photosphere ({\it e.g.}, \citealp{Hagyard1984, Leka1996}), inferred by applying Amp\`ere's law to the horizontal field (the components tangent to the photosphere). Although they have historically been rare, vector magnetograms have recently been made more frequently, thanks to NSO's SOLIS vector spectromagnetograph (VSM; \citealp{Keller2003}), the SpectroPolarimeter (SP; \citealp{Lites2013}) on the {\it Solar Optical Telescope} (SOT; \citealp{Tsuneta2008}) aboard the {\em Hinode} satellite \citep{Kosugi2007}, and the {\it Helioseismic and Magnetic Imager} (HMI; \citealp{Scherrer2012, Schou2012}) instrument aboard the {\it Solar Dynamics Observatory}.

For a given distribution of radial magnetic flux at the photosphere,
the hypothetical, curl-free coronal magnetic field matching this
distribution has the lowest magnetic energy among the set of coronal
fields consistent with the specified radial field ({\it e.g.},
\citealp{Priest2014}).
% p.117 of Priest 2014
Since this minimum-energy field is curl-free, by Amp\`ere's law it is
also current-free. Hence, the magnetic energy stored in coronal
electric currents is the energy above the minimum-energy state, and is
referred to as free magnetic energy ({\it e.g.}, \citealp{Forbes2000,
  Welsch2006}). This energy is available to be released impulsively in
events like flares and CMEs {\it via} dissipation of coronal
currents. Increases in free magnetic energy are therefore thought to
indicate a greater likelihood of flares and CMEs. Since the actual
coronal field $\vect{B}$ throughout the coronal volume cannot be
directly measured, its energy is usually estimated by some form of
modeling. Approaches include extrapolation based upon a vector
magnetogram (e.g., \citealt{DeRosa2009}) or modeling departures from
the energy of an initial, current-free state inferred from
photospheric magnetic evolution (see, e.g., \citealt{Kazachenko2015}
or \citealt{Cheung2012}).

Photospheric vector magnetic field observations therefore play a key role in estimating the coronal free energy present. Quantifying free energy requires estimating both the energy in the actual coronal magnetic field, $\vect{B}$, and the energy in the corresponding minimum-energy field. Since the latter is curl-free, it can be represented as the gradient of a scalar potential, $\vect{B}^{\rm P} = -\grad \chi$. This field is therefore often referred to as a potential field, hence the superscript P. The divergence-free condition on magnetic fields implies that $\chi$ obeys Laplace's equation, meaning that it is completely determined by boundary conditions (BCs) on the potential function. One goal of this paper is to investigate BCs on $\chi$ derived from vector magnetograms. \citet{Low1990} argued that trying to exactly match both the normal- and horizontal-field BCs was fundamentally flawed, and suggested using the photospheric normal field as a BC for potential field extrapolations. We believe, however, that extra information from observed horizontal fields can be usefully included in the potential field solution. \citet{Fisher2010} suggested that solutions which closely match both the normal field and the curl-free component of the horizontal field could be used to determine potential field structure in the neighborhood of the boundary. We present an approach for determining a potential field that matches both the normal field and the curl-free part of the horizontal field in a statistical sense, and can be used to extrapolate a potential field throughout the coronal volume.

In addition, while currents normal to the atmospheric layer imaged in vector magnetograms have been studied, few observational constraints have been placed on the presence or structure of horizontal currents within and near the photosphere. A second goal of this paper is to quantify differences between the observed and potential magnetic fields that indicate the presence of these horizontal currents. Theoretical considerations ({\it e.g.}, \citealp{Spruit1981}) imply that approximately solenoidal ``sheath'' currents should be present around the peripheries of active region (AR) flux systems in the solar interior, since the coherent magnetic fields that form active regions appear isolated from surrounding plasma that typically lacks any spatially coherent field. Sheath currents might also be present at the photosphere. (Although strong fields are present in ``quiet'' regions of the photosphere, their spatial structure is highly intermittent; see, {\it e.g.}, \citep{SanchezAlmeida2009}.)

The remainder of this paper is organized as follows. In Section \ref{sec:canon}, we discuss Neumann and Dirichlet BCs for extrapolating potential fields. In Section \ref{sec:hybrid}, we discuss two approaches for finding potential fields that borrow from both BCs: the first weights the normal field and curl-free horizontal field equally; the second, recognizing that uncertainties generally differ between the measured field components, uses weighting by parametrized noise estimates to explicitly incorporate uncertainties into the field extrapolation. Section \ref{sec:physics} explores some implications of our potential field models. We conclude with a brief summary of our main results, and a discussion of their implications.

\section{Canonical Potential Fields}
\label{sec:canon}
On the scale of a typical solar active region, one can approximate the spherical solar photosphere as a Cartesian plane. Here, we adopt this approach, defining $\hatz$ to be in the normal direction, with $z=0$ at the photosphere. In the three-dimensional (3D), Cartesian, half-space satisfying $z > 0$, the potential field $\vect{B}^{\rm P} = -\grad \chi$, where the scalar potential $\chi$ satisfies
\be \chi(x,y,z) = \frac{1}{(2\pi)^2}
\int_{-\infty}^{+\infty} {\rm d}k_x \int_{-\infty}^{+\infty} {\rm d}k_y \,
\tilde \chi(\vect{k}) {\rm e}^{{\rm i}k_xx + {\rm i}k_yy - k_zz}
~, \label{eqn:chidef} \ee
solves Laplace's equation if $k_z = \sqrt{k_x^2 + k_y^2} = k_h$, with the outer BC set by $|\vect{B}^{\rm P}| \to 0$ at infinity. This functional form implies that the spectral function $\tilde \chi(\vect{k})$ is a function of only the two independent wave numbers $k_x$ and $k_y$: the variation with height $z$ is determined by the spatial variations of the potential function in the $z=0$ plane. Typically either Neumann or Dirichlet BCs are imposed at $z=0$ to determine $\tilde \chi(\vect{k})$.

For the Neumann condition, the solution for the potential $\chi_{\rm N}$ on the $z=0$ boundary is then found from
\be B_z(x,y,0) = -\partial_z \chi_{\rm N} \vert_{z=0} =
\frac{1}{(2\pi)^2} \int_{-\infty}^{+\infty} {\rm d}k_x \int_{-\infty}^{+\infty} {\rm d}k_y \,
\tilde \chi_{\rm N} (k_x,k_y) k_h e^{{\rm i}k_xx + {\rm i}k_yy}
~, \label{eqn:neumann_bc} \ee
with
\be \tilde \chi_{\rm N} (k_x,k_y)
= \frac{
\int_{-\infty}^{+\infty} {\rm d}x
\int_{-\infty}^{+\infty} {\rm d}y \, B_z(x,y,0) {\rm e}^{-{\rm i}k_xx - {\rm i}k_yy} }{k_h}
= \frac{\tilde B_z(k_x,k_y)}
{k_h}
~, \label{eqn:neumann_chi} \ee
where we have defined $\tilde B_z(k_x,k_y)$ as the Fourier transform of $B_z(x,y)$. The spectral function determined by Equation (\ref{eqn:neumann_chi}) can then be used in Equation (\ref{eqn:chidef}) to determine the potential field for $z>0$ that satisfies the Neumann BC.

We now turn to the Dirichlet BC. In principle, knowledge of just one component of the measured horizontal field could be used to specify a Dirichlet BC: the curl-free condition on $\vect{B}_h$ implies $k_x \tilde B_y = k_y \tilde B_x$, where $\tilde B_y$ and $\tilde B_x$ are the Fourier transforms of $B_y$ and $B_x$, respectively. This could be used to determine the 2D potential function on $z=0$, a Dirichlet condition for the 3D Laplace's equation on $z > 0$ (see below). Such an approach would, however, ignore observations of the other horizontal component, and would therefore generally be inconsistent with these. In addition, this approach would also treat the data as exact, without accounting for the presence of uncertainties in the measurements.

Information from both components of the horizontal field are combined in computing the horizontal divergence of $\vect{B}$ at the photosphere, which gives
\be \divgh \vect{B}_h = -(\partial_x^2 \chi_{\rm D} + \partial_y^2 \chi_{\rm D} )
~. \label{eqn:diri_source} \ee
This is a 2D Poisson equation for $\chi_{\rm D}(x,y)$ at $z=0$. Specifying Dirichlet or Neumann BCs on $\chi_{\rm D}$ for the $x$ and $y$ boundaries (possibly at infinity) uniquely determines $\chi_{\rm D}$ on $z=0$, to within a constant \citep{Jackson1975}. Since $\vect{B}^{\rm P}$ only depends on derivatives of the potential function, this forms a Dirichlet condition for the 3D Laplace's equation for $\chi_{\rm D}$ for $z > 0$. The solution $\chi_{\rm D}$ on $z = 0$ is then found from
\bea \divgh \vect{B}_h(x,y,0)
&=& -(\partial_x^2 \chi_{\rm D} + \partial_y^2 \chi_{\rm D} ) \vert_{z=0} \\
&=& \frac{1}{(2\pi)^2} \int_{-\infty}^{+\infty} {\rm d}k_x \int_{-\infty}^{+\infty}
{\rm d}k_y \, \tilde \chi_{\rm D}(k_x,k_y) (k_x^2 + k_y^2) {\rm e}^{{\rm i}k_xx + {\rm i}k_yy}
~, \label{eqn:dirichlet_bc} \eea
with
\be \tilde \chi_{\rm D}(k_x,k_y)
= \frac{\int_{-\infty}^{+\infty} {\rm d}x \int_{-\infty}^{+\infty} {\rm d}y
[\divgh \vect{B}_h(x,y,0)] {\rm e}^{-{\rm i}k_xx - {\rm i}k_yy}}{k_h^2}
= \frac{{\rm i} \vect{k}_h \cdot \tilde{\vect{B}}_h(k_x,k_y)}{k_h^2}
~, \label{eqn:dirichlet_chi} \ee
where ${\rm i}\vect{k}_h \cdot \tilde{\vect{B}}_h(k_x,k_y)$ is the Fourier transform of $\divgh \vect{B}_h(x,y,0)$. (The divergence-free condition on magnetic fields implies that $|\vect{B}_h(x,y,0)|$ decays sufficiently fast with $x$ and $y$ that this integral converges for a localized source, such as an active region.) This spectral function $\chi_{\rm D}$ on $z=0$ can then be used with Equation (\ref{eqn:chidef}) to determine the potential field for $z>0$ that satisfies the Dirichlet BC.

Vector magnetograms can therefore be used to determine spectral functions for the potential field in at least two ways: using the Neumann BC to derive $\tilde \chi_{\rm N}$ from $B_z,$ and using the Dirichlet BC to derive $\tilde \chi_{\rm D}$ from $\divgh \vect{B}_h$, using Equations (\ref{eqn:neumann_chi}) and (\ref{eqn:dirichlet_chi}), respectively. If the magnetic field on the $z=0$ boundary is potential, then the uniqueness of solutions to Laplace's equation \citep{Jackson1975} implies that $\grad \tilde \chi_{\rm D}(k_x, k_y) = \grad \tilde \chi_{\rm N}(k_x,k_y)$ and therefore $B_z^{\rm P}$ and $\divgh \vect{B}_h^{\rm P} = -\partial_z B_z^{\rm P}$ are consistent. Even if the observed field $\vect{B}^{\rm obs}$ were potential, however, the presence of noise or systematic errors in the data would, in general, introduce inconsistencies between $\tilde \chi_{\rm N}$ and $\tilde \chi_{\rm D}$ for those observations. If, in addition, $\vect{B}^{\rm obs}$ is not potential, then generally no potential field will be simultaneously consistent with $\divgh \vect{B}_h^{\rm obs}$ and $B_z^{\rm obs}$. Differences between $\tilde \chi_{\rm N}$ and $\tilde \chi_{\rm D}$ might be useful to infer properties of either (i) noise / errors in the measurements or (ii) currents, or both.

We note that although our expressions for the Neumann and Dirichlet
solutions are only appropriate for a particular domain (the Cartesian,
$z > 0$ half space), a different choice of domain would not affect the
general conclusion: the observed normal field $B_n$ and $\divgh \vect
B_h$ at the photosphere separately determine Neumann and Dirichlet
BCs, respectively, for a solution $\chi$ to Laplace's equation in the
coronal domain, and the potential magnetic fields derived from each of
these BCs will, in general, differ due to measurement errors and
non-potentiality at the surface.

We implemented codes to rapidly compute a potential field from either Neumann or Dirichlet boundary conditions on Cartesian domains using discrete Fourier transforms (DFTs). We verified the accuracy of our reconstructed potential field (and its associated magnetic energy) using synthetic photospheric magnetic field data from a test case with a relatively easy analytic solution: the potential field from a submerged, vertically-oriented dipole.

To demonstrate that differences in the Neumann and Dirichlet potential
fields are present in solar data, we solve Laplace's equation for each
BC for a SDO/HMI vector magnetogram of AR 11158 from the sequence
analyzed by \citet{Welsch2013}. We arbitrarily chose an observation
from 15 February 2011, with the center of the integration time at
02:00 UT (near the end of an X2.2 flare), when the center of the
active region was at S20 W13. In images of the active region's fields,
we show a $426 \times 276$ pixel region containing essentially all of
the region's flux; the full magnetogram is $612 \times 610$ pixels,
and the area outside of the central region lacks any significant
spatially coherent magnetic field. These data have been interpolated
onto a Cartesian plane using a Mercator projection \citep{Welsch2009},
with a pixel size of 362 km. The ratio of total signed flux to the
total unsigned flux over the magnetogram is $< 1$\%.

\begin{figure}[ht]
\centerline{\includegraphics[width=15.5cm]{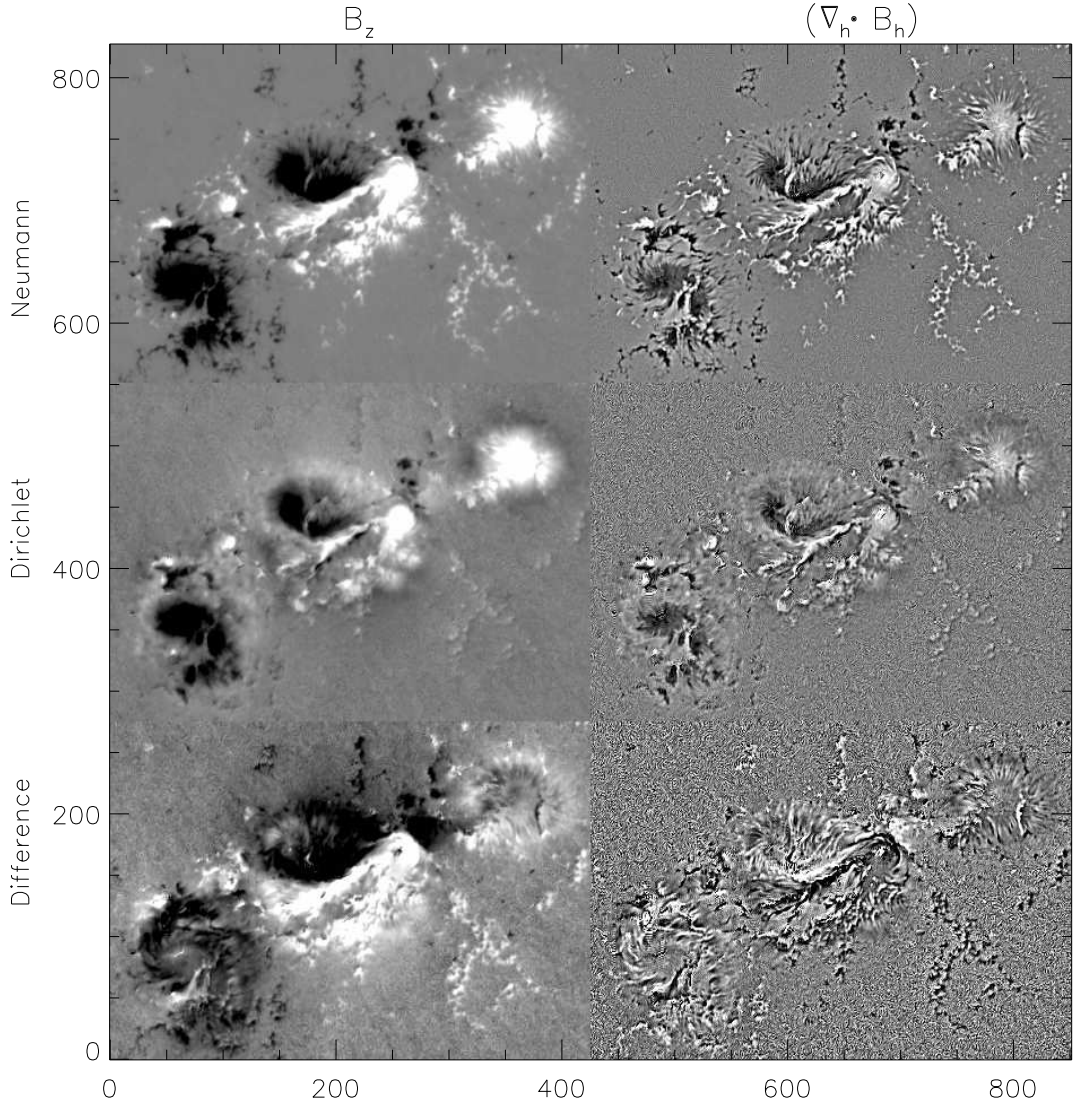}}
\caption{
Top row: Observed $B_z$ (left) was used as a Neumann condition to
derive $\divgh \vect{B}_h$ expected from a potential field
(right). Middle row: $B_z$ (left) for the potential field inferred
using the observed $\divgh \vect{B}_h$ (right) as a Dirichlet
BC. Bottom left: Difference between observed and reconstructed $B_z$
(top left minus middle left). Bottom right: Difference between
observed and reconstructed $\divgh \vect{B}_h$ (middle right minus top
right). The $x$ and $y$ axes are in interpolated, reprojected HMI
pixels, which are square and 362 km on a side. In the left column,
image saturation is (top to bottom): $\pm 1000$ Mx cm$^{-2}$, $\pm
1000$ Mx cm$^{-2}$, and $\pm 500$ Mx cm$^{-2}$. In the right column,
image saturation is (top to bottom): $\pm 333$ Mx cm$^{-2}$
pix$^{-1}$, $\pm 333$ Mx cm$^{-2}$ pix$^{-1}$, and $\pm 167$ Mx
cm$^{-2}$ pix$^{-1}$. Note the opposite-polarity halos around
strong-field regions in the image of $B_z$ derived from the Dirichlet
condition (middle-left panel).  }
\label{fig:bpot_comp}
\end{figure}

Figure \ref{fig:bpot_comp} compares the observed $B_z$ and $\divgh
\vect{B}_h$, and the versions of these reconstructed from the Dirichlet
and Neumann solutions, respectively, as well as the difference between
observed and reconstructed quantities. Note that the saturation levels
in this figure differ between left and right columns, and are smaller
in the bottom row. We have expressed the inverse length scale in
$\divgh \vect{B}_h$ as pix$^{-1}$ to facilitate comparisons with the
magnitude of $B_z.$ Note the opposite-polarity halos present around
strong-field regions in the middle-left panel, which shows $B_z$
derived from the Dirichlet BC. This systematic feature of
Dirichlet-derived normal potential fields cannot plausibly have arisen
from random errors in the magnetogram measurements, although they
might be due to systematic errors in the inversion process (discussed
further below). We also verified that these halos are not an artifact
of periodicity inherent in our use of DFTs to solve for the
potentials, by padding the $621 \times 610$ pixel magnetogram arrays
with a moat of 400 zeros. (The 426 $\times$ 276 arrays shown in this
and other figures were cropped close to the active region to display
its fields more clearly.) The vertical components of the Dirichlet
potential fields derived from the unpadded and padded magnetograms
closely match: the mean and maximum unsigned difference between the
two are 2 and 7 Mx cm$^{-2}$ (Mx: maxwell), respectively, over the 426
$\times$ 276 field of view shown here. Further, the halos, which are
coherent over tens of pixels, do not exhibit the high spatial
frequencies of Fourier ringing / Gibbs phenomena. (See also Figure
\ref{fig:bz_fe} in Appendix \ref{sec:dirich_tile}, which shows halos
also present in $B_z$ in a Dirichlet potential field computed using a
non-periodic, finite-element approach.) Assuming the $B_z$-halos reflect a
real property of $\divgh \vect{B}_h$, they indicate that the observed
vertical field is not consistent with the Dirichlet potential field
--- {\it i.e.}, the lack of halos in the observed $B_z$ indicates that
horizontal currents are present (also discussed further below).

\section{Hybrid Potentials: Using Both $B_z$ and $\divgh \vect{B}_h$}
\label{sec:hybrid}
As a starting point, we assume that the potential field should be the
current-free field that is, in some sense, {\em most consistent with
  observations.} As we have seen, the Neumann and Dirichlet solutions
are each consistent with distinct components of the observed
photospheric magnetic field. We now consider generalized definitions
of ``most consistent with observations.''

\subsection{Least-Squares Difference from the Data}
\label{subsec:hybrid}
Instead of computing a potential field that matches either the vertical or horizontal parts of the observed photospheric magnetic field, another approach is to determine a single potential field that is most consistent with all three components of the observed magnetic vector. Adopting the sum of squared residuals between the photospheric potential and magnetogram vector fields as our measure of such consistency, we can determine a ``combined'' potential $\chi_{\rm c}$ whose gradient minimizes the corresponding functional $I$,
\be I = \int_{-\infty}^{+\infty} {\rm d}x \int_{-\infty}^{+\infty} {\rm d}y
\left ( \vect{B}^{\rm obs} + \grad \chi_{\rm c} \right )^2
~, \label{eqn:i_def} \ee
where all quantities are evaluated at $z=0$. As posed, this is a type of least-squares minimization problem, albeit non-local.

Given a variation $\chi'$ about the extremal potential $\chi_{\rm c}$ that minimizes $I$, where $\chi'$ also obeys Equation (\ref{eqn:chidef}), the first-order variation $\delta I$ must vanish for arbitrary $\chi'$. The first-order variation is given by
\bea \delta I
&=& 2 \int_{-\infty}^{+\infty} {\rm d}x \int_{-\infty}^{+\infty} {\rm d}y
\left [ \grad \chi' \cdot \left ( \vect{B}_h^{\rm obs} + \grad \chi_{\rm c} \right ) \right ] \\
&=& 2 \int_{-\infty}^{+\infty} {\rm d}x \int_{-\infty}^{+\infty} {\rm d}y
\left [ \partial_z \chi' \left ( B_z^{\rm obs} + \partial_z \chi_{\rm c} \right ) +
\grad_h \chi' \cdot \left ( \vect{B}_h^{\rm obs} + \grad_h \chi_{\rm c} \right ) \right ] \\
&=& 2 \int_{-\infty}^{+\infty} {\rm d}x \int_{-\infty}^{+\infty} {\rm d}y
\left [ \partial_z \chi' \left ( B_z^{\rm obs} + \partial_z \chi_{\rm c} \right ) -
\chi' \left ( \divgh \vect{B}_h^{\rm obs} + \nabla_h^2 \chi_{\rm c} \right ) \right ] \nn \\
&&+ \oint {\rm d}\ell
\, \chi' \hatn \cdot (\vect{B}_h + \grad_h \chi_{\rm c} ) ~. \eea
In the final expression, $\hatn$ points horizontally outward along the closed curve defining the integral. We assume this integral vanishes due to either the finite spatial extent of the observed field's horizontal divergence or constraints on the imposed variation $\chi'$ at the $x$ and $y$ boundaries.

We Fourier transform $\chi'$ to a function of $\vect{k}_h'$, writing its spectral function as $\tilde \chi'$, giving
\bea \delta I
&=& \frac{1}{2 \pi^2} \int_{-\infty}^{+\infty} {\rm d}x \int_{-\infty}^{+\infty} {\rm d}y
\int_{-\infty}^{+\infty} {\rm d}k_x' \int_{-\infty}^{+\infty} {\rm d}k_y' {\rm e}^{{\rm i}k_x'x + {\rm i}k_y'y} \times \nn \\
&& \left [ - \tilde \chi'(\vect{k}_h') k_h'
\left ( B_z^{\rm obs} + \partial_z \chi_{\rm c} \right ) -
\tilde \chi'(\vect{k}_h')
\left ( \divgh \vect{B}_h^{\rm obs} + \nabla_h^2 \chi_{\rm c} \right ) \right ]
~, \label{eqn:fft1} \eea
where the differential operator $\partial_z$ acting on $\chi'$, evaluated at $z=0$, has brought down a factor of $-k_z'=-k_h'$. Next, we Fourier transform $B_z^{\rm obs}$, $\chi_{\rm c}$, and $\vect{B}_h^{\rm obs}$ into functions of $\vect{k}_h$, as above using tildes to denote spectral functions, and then act on these with the differential operators $\partial_z$, $\grad_h$, and $\nabla_h^2$, which bring down factors of $-k_h$, ${\rm i} \vect{k}_h$, and $-k_h^2$, respectively, to give
\bea \delta I
&=& \frac{1}{8 \pi^4} \int_{-\infty}^{+\infty} {\rm d}x \int_{-\infty}^{+\infty} {\rm d}y
\int_{-\infty}^{+\infty} {\rm d}k_x' \int_{-\infty}^{+\infty} {\rm d}k_y' {\rm e}^{{\rm i}k_x'x + {\rm i}k_y'y}
\int_{-\infty}^{+\infty} {\rm d}k_x  \int_{-\infty}^{+\infty} {\rm d}k_y {\rm e}^{{\rm i}k_x x + {\rm i}k_y y}
 \times \nn \\
&& \left [ - \tilde \chi'(\vect{k}_h') k_h' \left \{ \tilde B_z^{\rm obs}(\vect{k}_h)
   - k_h \tilde \chi_{\rm c}(\vect{k}_h) \right \} -
\tilde \chi'(\vect{k}_h')
\left \{ {\rm i} \vect{k}_h \cdot \tilde{\vect{B}}_h^{\rm obs}(\vect{k}_h)
- k_h^2 \tilde \chi_{\rm c}(\vect{k}_h)
\right \} \right ] ~. \eea
We then interchange the spatial and wavenumber integrations, and note that
\be \frac{1}{2 \pi}
\int_{-\infty}^{+\infty} {\rm d}x \, {\rm e}^{{\rm i}(k_x + k_x') x} = \delta(k_x + k_x')
\label{eqn:delta_fcn} ~, \ee
and similarly for the ${\rm d}y$ integral. Hence, integrating over $x$ and $y$ implies $\vect{k}_h' = -\vect{k}_h$, so
\be \delta I
= \frac{-1}{2 \pi^2} \int_{-\infty}^{+\infty} {\rm d}k_x \int_{-\infty}^{+\infty} {\rm d}k_y
\, \tilde \chi'(-\vect{k}_h) \left [ k_h
\left \{ \tilde B_z^{\rm obs}(\vect{k}_h) - k_h \tilde \chi_{\rm c}(\vect{k}_h) \right \} +
{\rm i} \vect{k}_h \cdot \tilde{\vect{B}}_h^{\rm obs}(\vect{k}_h) - k_h^2 \tilde \chi_{\rm c}(\vect{k}_h)
 \right ] ~. \\
\ee
Since $\delta I$ must vanish for an arbitrary variation $\chi'$ about the extremal potential $\chi_{\rm c}$, the quantity in square brackets must be zero, {\it i.e.},
\be
0 = k_h
\left \{\tilde B_z^{\rm obs}(\vect{k}_h) - k_h \tilde \chi_{\rm c}(\vect{k}_h) \right \} +
{\rm i} \vect{k}_h \cdot \tilde{\vect{B}}_h^{\rm obs}(\vect{k}_h) - k_h^2 \tilde \chi_{\rm c}(\vect{k}_h) 
~, \label{eqn:constraint_0} \ee
which implies
\be
\tilde \chi_{\rm c}(\vect{k}_h) =
[k_h \tilde B_z^{\rm obs}(\vect{k}_h) +
{\rm i} \vect{k}_h \cdot \tilde{\vect{B}}_h^{\rm obs}(\vect{k}_h)]
/(2 k_h^2) ~. \label{eqn:combo_spect} \ee
Hence, this spectral function can be determined directly from observed quantities ({\it i.e.}, the spectral functions of $B_z^{\rm obs}$ and $\vect{B}_h^{\rm obs}$), from which this combined potential field can be calculated. Comparison of Equation (\ref{eqn:combo_spect}) with Equations (\ref{eqn:neumann_chi}) and (\ref{eqn:dirichlet_chi}) shows that the combined result is an average of the Neumann and Dirichlet spectral functions. This result might be expected, from the known least-squares properties of averaging, but it nicely confirms our earlier argument that the horizontal divergence is a good choice for specifying the Dirichlet BC, as in Equation (\ref{eqn:dirichlet_bc}).

\begin{figure}[ht]
\centerline{\includegraphics[width=15cm]{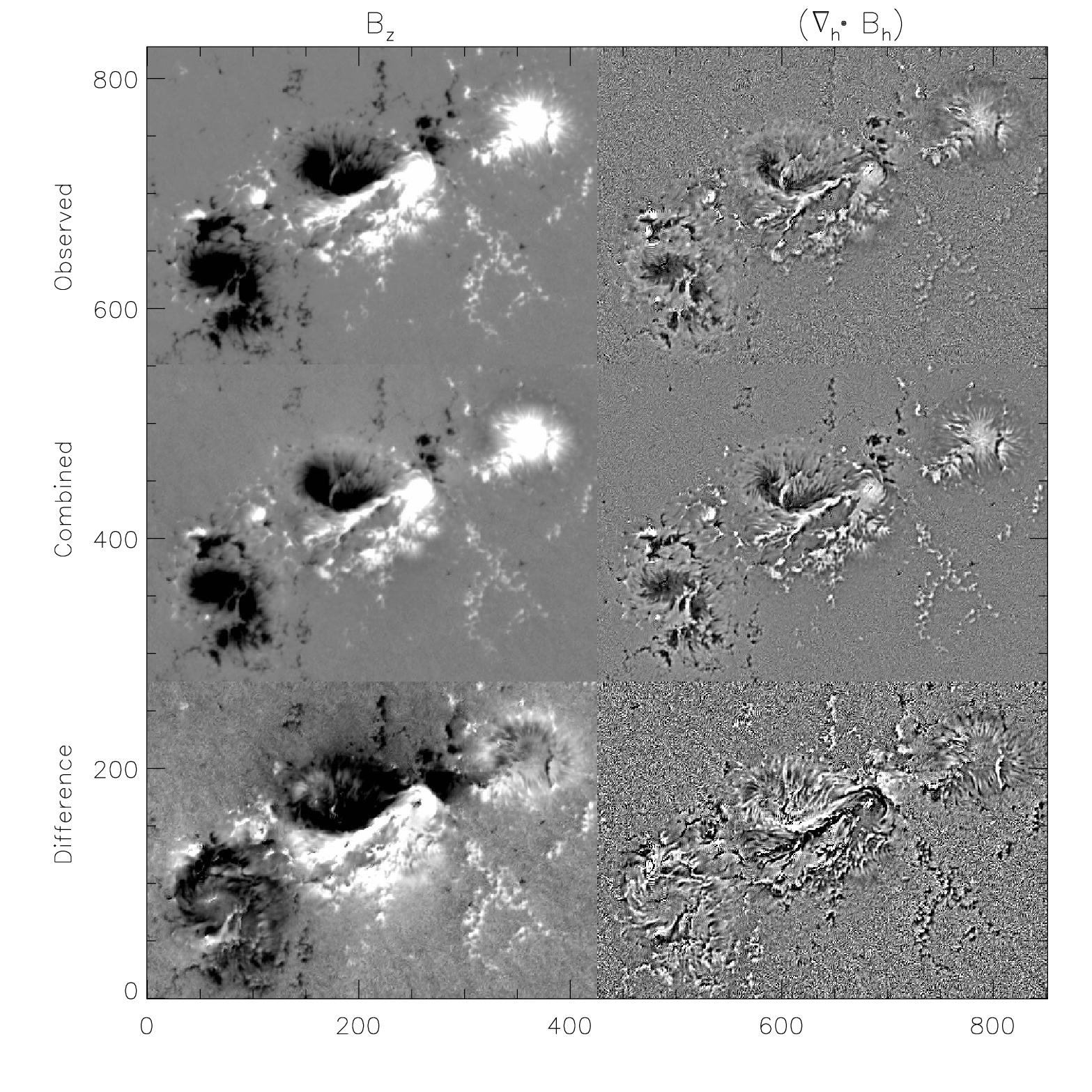}}
\caption{%
  Top row: Observed $B_z$ (left) and $\divgh \vect{B}_h$
  (right). Middle row: Combined potential field's $B_z$ (left) and
  $\divgh \vect{B}_h$ (right), derived from the spectral function in
  Equation (\ref{eqn:combo_spect}). Bottom left: Difference between
  observed and reconstructed $B_z$ (top left minus middle
  left). Bottom right: Difference between observed and reconstructed
  $\divgh \vect{B}_h$ (top right minus middle right).
The $x$ and $y$ axes are in interpolated, reprojected HMI pixels,
which are square and 362 km on a side.
In the left column, image saturation is (top to bottom): $\pm 1000$ Mx
cm$^{-2}$, $\pm 1000$ Mx cm$^{-2}$, and $\pm 250$ Mx cm$^{-2}$. In the
right column, image saturation is (top to bottom): $\pm 333$ Mx
cm$^{-2}$ pix$^{-1}$, $\pm 333$ Mx cm$^{-2}$ pix$^{-1}$, and $\pm 83$
Mx cm$^{-2}$ pix$^{-1}$.  }
\label{fig:combo_comp}
\end{figure}

We have used this combined spectral function with DFTs to compute the combined potential field for the vector magnetogram from Figure \ref{fig:bpot_comp}. To confirm that the combined field does agree more closely than either the Neumann or Dirichlet fields, we compare the sums of squared differences between each potential field and the observed field, by substituting each potential function in for $\chi$ in Equation (\ref{eqn:i_def}). We find $I_{\rm N} = 1.74 \times 10^{25}$ Mx$^2$ cm$^{-2}$, $I_{\rm D} = 1.77 \times 10^{25}$ Mx$^2$ cm$^{-2}$, and $I_{\rm c} = 1.36 \times 10^{25}$ Mx$^2$ cm$^{-2}$, for the Neumann, Dirichlet, and combined fields, respectively. In Figure \ref{fig:combo_comp}, we show the observed $B_z$ and $\divgh \vect{B}_h$ in the top row, reconstructions of these from the combined potential solution in the middle row, and the difference between observed and reconstructed quantities in the bottom row. Note that the image saturation in the bottom row of Figure \ref{fig:combo_comp} is set lower than that in Figure \ref{fig:bpot_comp}, reflecting the fact that the combined solution agrees more closely with the observations. In this figure, note that the opposite-polarity halos around strong-field regions in the image of $B_z$ from the combined field (middle-left panel) are discernible, but weaker than those the middle-left panel of Figure \ref{fig:bpot_comp}. We will discuss the physical significance of differences between potential and observed fields in more detail in Section \ref{sec:physics} below.

\begin{figure}[ht]
\centerline{\includegraphics[width=11cm]{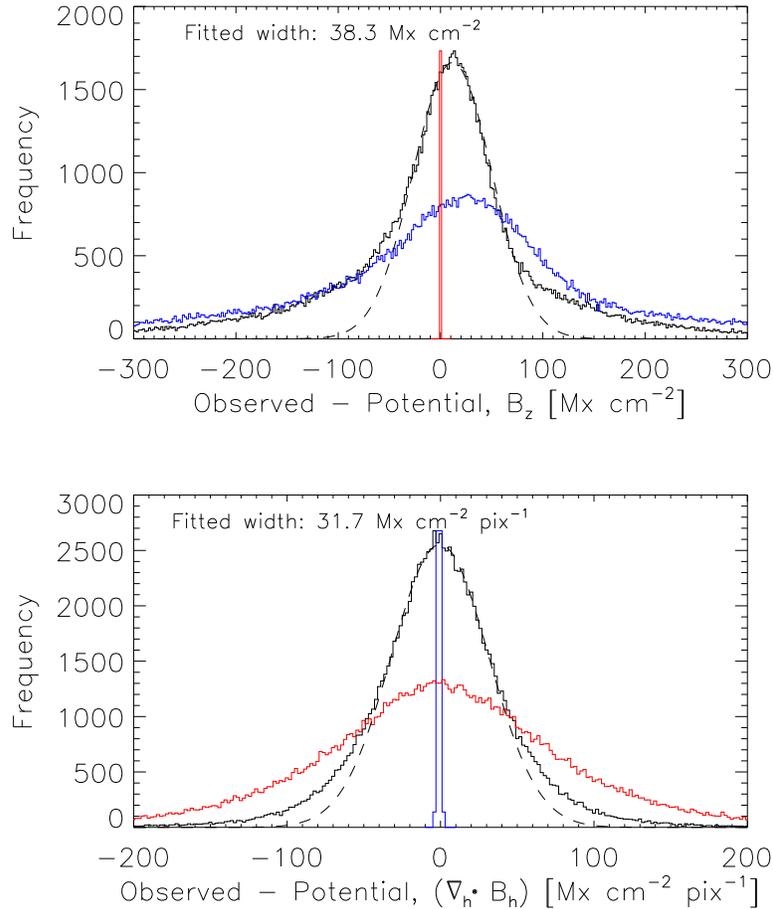}}
\caption{%
Top: Histograms of observed minus potential $B_z$, for Neumann (red), Dirichlet (blue), and combined (black solid) potential fields. The Neumann field's histogram has been rescaled to the maximum of the combined field's. Bottom: Analogous histograms of observed minus potential $\divgh \vect{B}_h$ for Neumann (red), Dirichlet (blue), and combined (solid black) fields. Here, the Dirichlet field's histogram has been rescaled to the maximum of the combined field's. As expected, the combined field's difference histograms fall between the other fields' histograms for each variable. In both cases, a Gaussian fit to the combined distribution is plotted with a black dashed line, and its fitted width is printed. The width is a measure of the significance of these field differences relative to noise levels in $B_z$ and $\divgh \vect{B}_h$; see text.
}
\label{fig:diff_hists}
\end{figure}

In the top panel of Figure \ref{fig:diff_hists}, we show histograms of differences in (signed) $B_z$ between the observations and each of these three potential fields --- Neumann (red), Dirichlet (blue), and combined (black solid). Differences are small in the Neumann case, since the Neumann field is constructed to match $B_z$; its histogram has been rescaled to the maximum of the combined field's. Differences are large, however, in the Dirichlet case. As expected, the combined field's differences fall between the other fields' differences. A Gaussian fit to the combined distribution is plotted with a black dashed line, and its fitted width is printed. This can be used to relate the magnitude of field differences to noise levels in the observed quantities. In the bottom panel, we show analogous histograms of differences in $\divgh \vect{B}_h$ between the observations and each of the three potential fields --- as above, Neumann is in red, Dirichlet in blue, and combined in solid black. In contrast to differences in $B_z$, now it is the Dirichlet case in which discrepancies are small (since the Dirichlet field is constructed to match $\divgh \vect{B}_h$), and its histogram has been rescaled to the maximum of the combined field's. As expected, differences from the Neumann case are large, and differences from the combined field fall between the two other two cases. Again, a Gaussian fit to differences from the combined field is overplotted with a black dashed line, and the fitted width is printed. (Our use of DFTs to solve for Dirichlet and Neumann fields yields flux-balanced fields to about 1 part in 10$^6$ or better.)

Equation (\ref{eqn:combo_spect}) determines the spectral function of
the potential field that satisfies the particular definition of ``most
consistent with observations'' --- the minimized, integrated, squared
discrepancy --- given in Equation (\ref{eqn:i_def}). Since any other
potential function (other than a trivial constant function) added to
$\chi_{\rm c}$ will alter the total potential and/or its normal
derivative on the boundary \citep{Jackson1975}, this least-squares
potential is unique. As we have seen, either the Neumann or Dirichlet
potential can closely match either vertical or horizontal components
of the photospheric field, respectively. But in a statistical,
least-squares sense, neither matches the total photospheric field,
over the full magnetogram, as closely as the combined solution. Also,
even where they do match the observations, the Neumann and Dirichlet
approaches have a clear drawback in that they overfit the data: the
noise certainly present in the fitted component of the observed field
is matched precisely (in principle, to machine round-off error), as
though it were real signal. This is an issue of precision versus
accuracy: either the Neumann or Dirichlet solution can match the
corresponding field component more precisely than the accuracy of its
measurement. In contrast, the combined potential field matches the
observations statistically.

\subsection{Uncertainty-Weighted Potential Fields}
\label{subsec:weighted}
Uncertainties in measurements of line-of-sight (LOS) fields are typically smaller than uncertainties in transverse fields ({\it e.g.}, \citealp{Hoeksema2014, Kazachenko2015}). Consequently, for magnetograms of active regions near disk center, where the LOS and photospheric normal directions coincide, one could argue that the vertical photospheric field is measured more accurately than the horizontal field, and, therefore, that the Neumann BC should be preferred over the Dirichlet condition.

Two counterarguments can be made. First, even if uncertainties near disk center are much smaller for $B_z$ than for $\divgh \vect{B}_h$, completely ignoring $\divgh \vect{B}_h$ is unwarranted: even if noisier, the horizontal field measurements do contain additional information about the structure of the photospheric magnetic field. Given that the observations are uncertain, overdetermining the solution by using all available data should yield a result more consistent with observations. Second, uncertainties in the Neumann condition increase away from disk center, since the measured normal field includes both LOS and transverse components. Hence, noise levels for the Neumann condition can become comparable to those for the Dirichlet condition. (We note, however, that deriving the Dirichlet condition involves differentiating the data, which will increase uncertainties in the BC: a four-point finite-difference stencil for the horizontal divergence effectively doubles the error in the horizontal divergence compared to the error in the horizontal field.)

These considerations raise the question: in the presence of differing noise levels in the vertical and horizontal fields, how should these data be used when computing the potential field? Intuitively, we expect the BC should involve both vertical and horizontal fields, but these should be weighted by their noise levels. Following the approach outlined above, we seek a ``noise-weighted potential'' $\chi_{\rm w}$ whose gradient minimizes a functional $I_{\rm w}$ equal to the integral of the {\em weighted} sum of the squared residuals between the potential field and the observed field, with differing weights $w_z$ and $w_h$ for $B_z$ and $B_h$, respectively,
\be I_{\rm w} = \int_{-\infty}^{+\infty} {\rm d}x \int_{-\infty}^{+\infty} {\rm d}y \left [
w_z ( B_z^{\rm obs} + \partial_z \chi_{\rm w} )^2 +
w_h ( \vect{B}_h^{\rm obs} + \grad_h \chi_{\rm w} )^2
\right ] ~. \label{eqn:iw_def} \ee
One can then seek conditions for which the first order variation $\delta I_{\rm w}$ vanishes for any variation $\chi'$ about the noise-weighted potential $\chi_{\rm w}$.

One might use weights $w_z$ and $w_h$ that vary in space, for instance to account for either geometric dependence of uncertainties in the LOS and transverse measurements as functions of disk position or uncertainties propagated through the field inversion process in each pixel. Potential magnetic fields are, however, non-local, so errors in the underlying field measurements will necessarily be propagated, perhaps globally: a poor measurement of vertical flux density in a pixel will affect the potential solution outside of that pixel, even if that particular pixel's field is not strongly weighted in the fit. As a practical matter, use of spatially varying weights complicates the analytic minimization of $I_{\rm w}$. While numerical approaches might enable determination of a potential field that minimizes $I_{\rm w}$ with spatially varying weights, for simplicity we assume $w_z$ and $w_h$ are constants here. The analysis leading to Equation (\ref{eqn:constraint_0}) is then essentially unchanged, and leads to the constraint equation on the spectral function $\tilde \chi_{\rm w}(\vect{k}_h)$
\be
0 = w_z k_h
\left[\tilde B_z^{\rm obs}(\vect{k}_h) - k_h \tilde \chi_w(\vect{k}_h) \right] +
w_h \left[ {\rm i} \vect{k}_h \cdot \tilde{\vect{B}}_h^{\rm obs}(\vect{k}_h)
- k_h^2 \tilde \chi_{\rm w}(\vect{k}_h) \right]
~, \label{eqn:constraint_w} \ee
which implies
\be
\tilde \chi_{\rm w}(\vect{k}_h) = \frac{1}{w_z + w_h}
\frac{[w_z k_h \tilde B_z^{\rm obs}(\vect{k}_h)
+ w_h {\rm i} \vect{k}_h \cdot \tilde{\vect{B}}_h^{\rm obs}(\vect{k}_h)]
}{k_h^2} ~. \label{eqn:weight_spect_0} \ee
If we set
\bea w_z &=& \frac{1}{\sigma_z} \\
w_h &=& \frac{1}{\sigma_h} ~, \eea
where $\sigma_z$ and $\sigma_h$ represent the uncertainties in $B_z$ and $\Delta s (\divgh \vect{B}_h)$, respectively, and $\Delta s$ is the width of pixels (assumed square), then
\be \tilde \chi_{\rm w}(\vect{k}_h) = \frac{1}{\sigma_z + \sigma_h}
  \frac{[\sigma_h k_h
      \tilde B_z^{\rm obs}(\vect{k}_h) + \sigma_z {\rm i} \vect{k}_h \cdot
      \tilde{\vect{B}}_h^{\rm obs}(\vect{k}_h)]}{k_h^2} ~. \label{eqn:weight_spect_1}
\ee

\begin{figure}[ht]
\centerline{\includegraphics[width=4.5in]{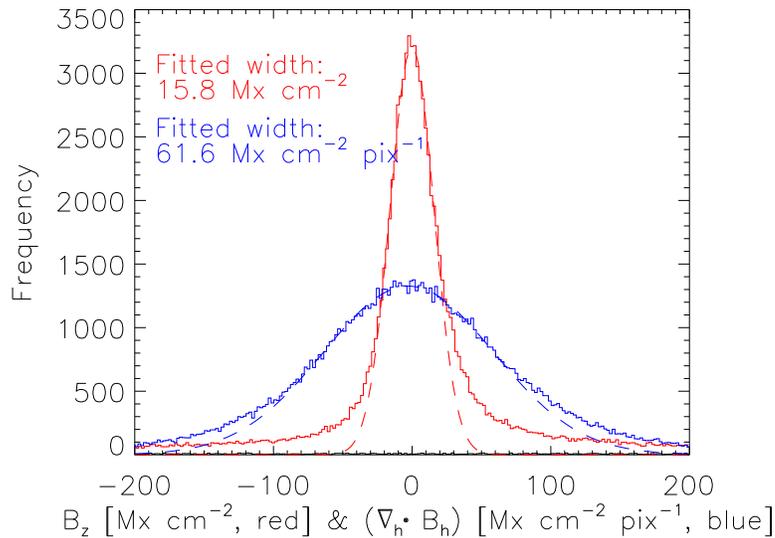}}
\caption{%
Histograms of observed $B_z$ (red solid) and $\divgh \vect{B}_h$ (blue solid). Gaussian fits to the core of each distribution (within $\pm$ 20 Mx cm$^{-2}$ and $\pm$ 60 Mx cm$^{-2}$ pix$^{-1}$, respectively) are overplotted as dashed lines in corresponding colors, and the widths of each fit are given. At this time, AR 11158 was at S20 W13.
}
\label{fig:bz_div_hists}
\end{figure}

To explore this weighting with real data, we estimated noise levels from histograms of $B_z$ and $\divgh \vect{B}_h$ from the SDO/HMI magnetogram, assuming the cores of each distribution arise from noise. We defined the cores to be values within $\pm$ 20 Mx cm$^{-2}$ and $\pm$ 60 Mx cm$^{-2}$ pix$^{-1}$, respectively, and fitted each core to a Gaussian. Figure \ref{fig:bz_div_hists} shows the distributions and fits. The fitted widths for $B_z$ and $\divgh \vect{B}_h$ are 15.8 Mx cm$^{-2}$ and 61.6 Mx cm$^{-2}$ pix$^{-1}$, respectively, which we interpret as the noise levels in each quantity. (Varying the widths of the fitted core regions did not change the fitted widths significantly.) This implies weights of about 4:1 on the Neumann versus Dirichlet condition.

\begin{figure}[ht]
\centerline{\includegraphics[width=4.5in]{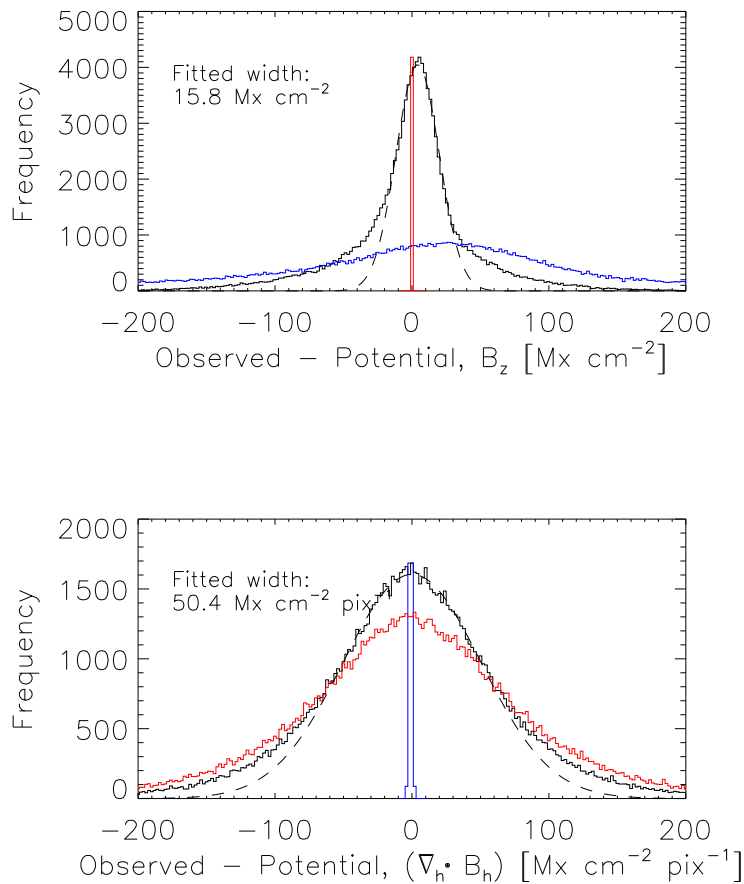}}
\caption{%
Top: Histograms of observed minus potential $B_z$, for Neumann (red), Dirichlet (blue), and noise-weighted (black solid) potential fields. The Neumann field's histogram has been rescaled to the maximum of the combined field's. Bottom: Analogous histograms of observed minus potential $\divgh \vect{B}_h$ for Neumann (red), Dirichlet (blue), and noise-weighted (solid black) fields. Here, the Dirichlet field's histogram has been rescaled to the maximum of the combined field's. A Gaussian fit to each noise-weighted distribution is plotted with a black dashed line, and its fitted width is printed. The widths of the noise-weighted distributions are commensurate with the noise levels in the observed quantities, shown in Figure \ref{fig:bz_div_hists}, used to determine the BC.
}
\label{fig:diff_hists_noise}
\end{figure}

We then used these noise levels in Equation (\ref{eqn:weight_spect_1}) to derive a noise-weighted potential $\chi_{\rm w}$ and its associated fields. Evaluating the sum of squared differences in Equation (\ref{eqn:i_def}) using $\chi_{\rm w}$ yields $I_{\rm w} = 1.56 \times 10^{25}$ Mx$^2$ cm$^{-2}$, so the weighted potential agrees more closely with the observed field than either the Neumann or Dirichlet fields, but does not agree as closely as the unweighted combined field. (As noted above, $I_{\rm N} = 1.74 \times 10^{25}$ Mx$^2$ cm$^{-2}$, $I_{\rm D} = 1.77 \times 10^{25}$ Mx$^2$ cm$^{-2}$, and $I_{\rm c} = 1.36 \times 10^{25}$ Mx$^2$ cm$^{-2}$.) This field's $B_z$ does not look significantly different than those from the combined field shown in Figure \ref{fig:combo_comp}, so we have not included a separate figure to show it. Histograms of the residuals, however, are different, as we show in Figure \ref{fig:diff_hists_noise}. It can be seen that the relative weighting has, as expected, improved consistency with the observed $B_z$, at the expense of decreased consistency with the observed $\divgh \vect{B}_h$. For each observable, the fitted widths of the cores of the distributions ($\pm 20$ Mx cm$^{-2}$ for $B_z$, $\pm 60$ Mx cm$^{-2}$ pix$^{-1}$ for $\divgh \vect{B}_h$) of the residuals are similar to our estimated noise levels. We note that the weights of each observable could be changed to vary the widths of the residuals' distributions. Since typical weighted least-squares fits usually weight by the {\em square} of the uncertainties in the fitted variable, we also tried this weighting. Given the roughly 1:4 ratio of noise levels between the vertical and horizontal fields, respectively, this gives a weighting of 1:16, heavily favoring the Neumann solution. With this ``squared noise weighting'' (snw), the sum of squared differences from Equation (\ref{eqn:i_def}) yields $I_{\rm snw} = 1.68 \times 10^{25}$ Mx$^2$ cm$^{-2}$, worse overall than weighting by the first power of each uncertainty.

\begin{figure}[ht]
\centerline{\includegraphics[width=4.25in]{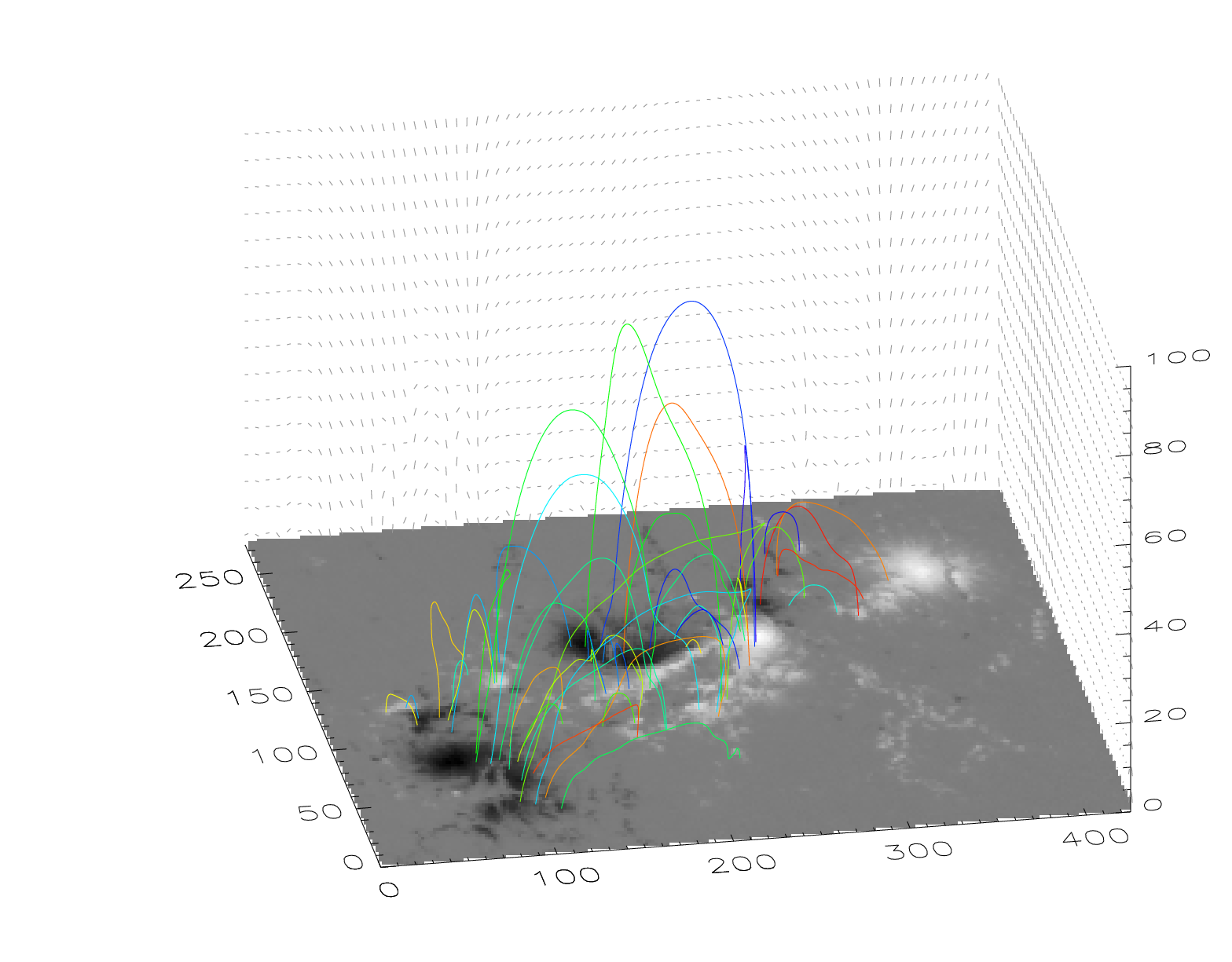}}  % 3.25in for side-by-side
\centerline{\includegraphics[width=4.25in]{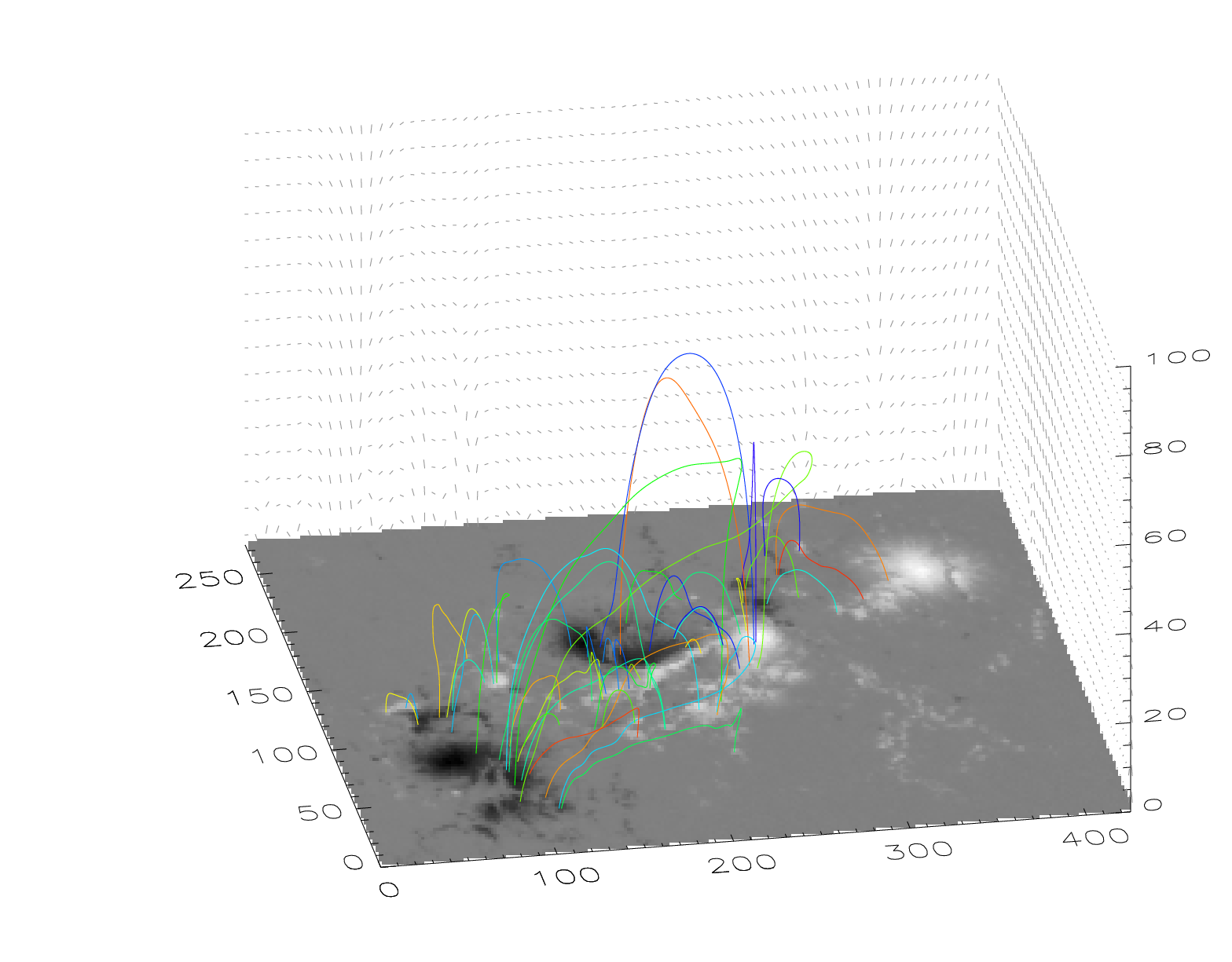}}
\caption{%
  Top: selected field lines integrated in the Neumann
  potential field. Bottom: selected field lines integrated in the
  noise-weighted potential field. Integrations were initiated from the
  same footpoints, and are plotted with the same color.
The $x$ and $y$ axes are in interpolated, reprojected HMI pixels,
which are square and 362 km on a side.
All field lines shown have both endpoints at the base of box.
}
\label{fig:field_lines}
\end{figure}

How do the extrapolated fields differ, qualitatively? A selection of field lines is plotted in Figure \ref{fig:field_lines}, for both the Neumann and noise-weighted hybrid potential fields. In the figure, field line integrations were initialized from the same footpoints with the same colors, but only field lines with both footpoints anchored at the photosphere are plotted. Many field lines look essentially the same, but some differences can be discerned. We note that the periodicity implicit in the DFT method used to compute these potential fields can affect field line connectivities. Minor differences aside, the fields' structures appear similar. Since Neumann potential coronal fields in some cases agree qualitatively with observed coronal loops (see, {\it e.g.}, \citealp{Schrijver2005}), it is reassuring that the noise-weighted coronal field exhibits similar morphology. We remark that observations of coronal magnetic field structure cannot readily be used to determine whether one approach to potential field extrapolation is ``better'' than another, since the coronal field is presumed to not be potential. We defer detailed investigation of differences in the fields' connectivities to a future study.

We have minimized a weighted sum of squared residuals, in a manner analogous to the weighting used in least-squares fitting. As with the unweighted minimization, for a fixed choice of weights, the solution to this minimization is unique. We separately weighted the horizontal- and vertical-field residuals by our distinct, constant estimates of the uncertainties in each.

\section{Physical Significance of Hybrid Boundary Conditions}
\label{sec:physics}
\subsection{Implications for Free Energy Estimates}
\label{subsec:energy}
The free magnetic energy in the coronal magnetic field is the difference between the energy in the actual coronal field and the energy of a hypothetical, minimum-energy field. It can be shown that the Neumann potential field is the lowest-energy field consistent with a given normal magnetic field at the photosphere ({\it e.g.}, \citealp{Priest2014}).

While $\vect{B}$ cannot currently be measured throughout the coronal volume above an active region, departures from the potential state can be estimated by either modeling $\vect{B}$ in the corona ({\it e.g.}, \citealp{DeRosa2009, Cheung2012}), or quantifying the photospheric Poynting flux ({\it e.g.}, \citealp{Kazachenko2015}). Quantifying coronal free energy is a key goal of such efforts.

Do different choices of BCs for potential fields result in different
energies for the potential field? Simply put: yes. The energy $U$ of a
potential field can be written as a surface integral of the potential
function times the normal magnetic field ({\it e.g.},
\citealp{Welsch2006}) at the photosphere,
\be U = \frac{1}{8\pi} \int {\rm d}A \, B_z \, \chi
~, \label{eqn:upot} \ee
where $B_z$ is the normal field that is consistent with the potential
function $\chi$, so $B_z$ in Equation (\ref{eqn:upot}) differs from
$B_z^{\rm obs}$ for all potentials but the Neumann solution.
Using the observed vector magnetic fields shown in the top row of
Figure \ref{fig:bpot_comp}, the energies of the Dirichlet, Neumann,
combined, and noise-weighted potential fields (derived using the
spectral functions in equations \ref{eqn:neumann_chi},
\ref{eqn:dirichlet_chi}, \ref{eqn:combo_spect}, and
\ref{eqn:weight_spect_1}, respectively) are $ 4.6 \times 10^{32}, 9.1
\times 10^{32}, 6.1 \times 10^{32},$ and $7.7 \times 10^{32}$ erg,
respectively. Because the Dirichlet field's energy is roughly a factor
of two lower than the Neumann field's, the combined and noise-weighted
potential-field energies are also significantly smaller than the
Neumann potential-field energy. It seems plausible that the excess
energy in the Neumann solution arises because the observed normal
field itself is, in a key sense, non-potential. This is consistent
with observations of flare-related changes to photospheric magnetic
fields, which in many cases show fields becoming ``more horizontal''
({\it e.g.}, \citealp{Wang2010}). We revisit this point below.

Given the relative noise levels on the normal and horizontal fields
for this magnetogram, it could be argued that the Dirichlet and
combined fields are too inconsistent with the observed normal field,
so discrepancies in their field energies are unimportant. The
noise-weighted potential field, however, is statistically consistent
with the observed normal field, but the energy in the Neumann
potential field is more than $10^{32}$ erg larger. This discrepancy is
a significant amount of energy, commensurate with that released in
large flares that produce fast CMEs ({\it e.g.},
\citealp{Emslie2012}). Evidently, the choice of BCs could have
significant consequences for modelers attempting to determine free
energies in coronal fields.

It should be noted, however, that our use of DFTs can produce
potential energies systematically larger than other methods ({\it
  e.g.}, a Green's function approach; \citealp{Sakurai1982}). This is
because our AR is implicitly assumed to be surrounded by neighboring
copies of itself, meaning its field is partially confined by the
magnetic pressure from its hypothetical neighbors. This can produce a
volume-averaged $B^2$ higher than that for an isolated AR. But this
bias should not affect the {\em ordering} in energies yielded by each
approach.

Based upon Equation (\ref{eqn:upot}), it is plausible that the energy
differences between the different potential fields might be related to
differences in the amount of unsigned vertical flux in each potential
model. Accordingly, we summed the unsigned vertical magnetic flux in
each model over all pixels with $|B_z| > 100$ Mx cm$^{-2}$, which gave
$2.3 \times 10^{22}$ Mx for the Dirichlet field, $2.8 \times 10^{22}$
Mx for the Neumann field, and $2.2 \times 10^{22}$ Mx for the combined
field. This threshold was chosen to avoid including $|B_z|$ from
weak-field regions in the Dirichlet solution, caused by spurious
values of $\divgh \vect{B}_h$ from differentiating noisy data
there. With this threshold, the Dirichlet summation included about
$5.3 \times 10^4$ pixels, the Neumann summation included about $3.8
\times 10^4$ pixels, and the combined solution included about $3.3
\times 10^4$ pixels. The larger number of pixels in the Dirichlet sum
probably reflects greater contributions from noise relative to the
sums from other fields. With this caveat in mind, we note that the
ordering of field energies does not follow the ordering of fluxes in
each model. The differences between the observed $B_z$ and the
Dirichlet field's $B_z$ shown in the lower-left panel of Figure
\ref{fig:bpot_comp} imply that the Dirichlet solution's vertical field
is systematically weaker in regions where the observed vertical field
is strong.  This, we believe, explains why the Dirichlet solution has
less unsigned magnetic flux than the Neumann solution, despite the
$B_z$-halos in the former.

\begin{figure}[ht]
\centerline{\includegraphics[width=4.5in]{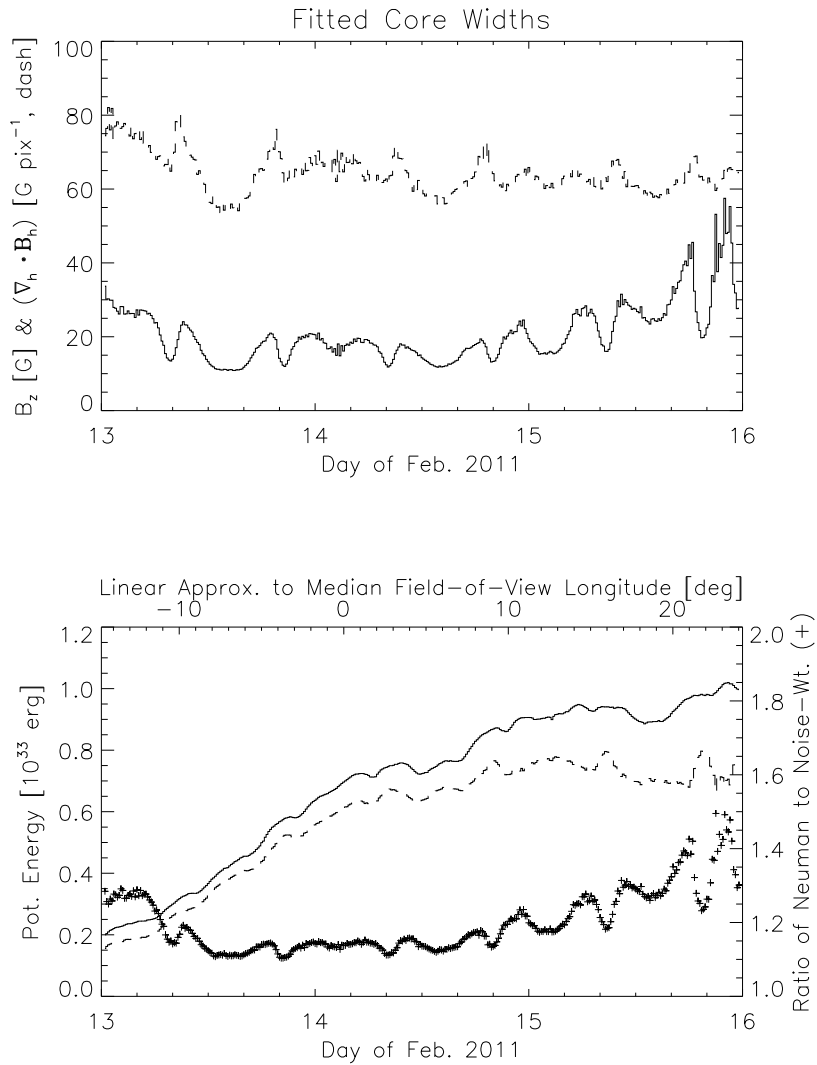}}
\caption{%
Top: Widths from Gaussian fits to cores of histograms of $B_z$ (solid) and $\divgh \vect{B}_h$ (dashed; cores are $\pm 20$ Mx cm$^{-2}$ and $\pm 60$ Mx cm$^{-2}$ pix$^{-1}$, resp.) versus time for three days of vector SDO/HMI magnetograms. (Vertical axis uses G instead of Mx cm$^{-2}$ for readability.) Bottom: Magnetic energies (left axis) in Neumann (solid) and noise-weighted hybrid (dashed) potential fields, and ratio of energies (+'s, right axis). A linear approximation to the median longitude of the active region's pixels is plotted on the top axis of the bottom plot; the active region's latitude remained about S20 throughout. We assume core widths are measures of uncertainty, and widths for $B_z$ (top plot, solid) increase toward the limb. The noise-weighted potential field incorporates more of the Dirichlet information toward the limb, so the ratio of energies increases.
}
\label{fig:sigmas_energies}
\end{figure}

In the top panel of Figure \ref{fig:sigmas_energies}, we show, as
functions of time, the widths from Gaussian fits to the cores of
histograms of $B_z$ and $\divgh \vect{B}_h$, where the core regions are
defined to be within $\pm 20$ Mx cm$^{-2}$ and $\pm 60$ Mx cm$^{-2}$
pix$^{-1}$, respectively. A linear approximation to the median
longitude of the active region's pixels is plotted on the top axis of
the bottom plot. (The active region's latitude remained about S20
throughout.) We assume core widths measure uncertainties in $B_z$ and
$\divgh \vect{B}_h$, and fitted widths increase toward the limb, as
expected. The bottom panel shows magnetic energies (left axis) in the
Neumann and noise-weighted hybrid potential fields, and their ratio
(right axis). Since noise in the Neumann condition increases toward
the limb, the noise-weighted potential field incorporates more of the
Dirichlet information toward the limb; and since the Dirichlet energy
is systematically lower than the Neumann energy, the ratio of Neumann
to noise-weighted energies increases. Over most of the interval
plotted, the energy of the Neumann potential field is 10\% (or more)
higher than that of the noise-weighted field, implying estimates of
free energy would be about 10\% (or more) lower if the Neumann field
is defined as the minimum energy state. Given the variation in weights
versus disk position in the noise-weighted approach, we favor making
comparisons between potential energies at different disk positions
determined using fixed weights, as opposed to the position-dependent
weights here.

\subsection{Do Other Magnetographs Show Similar Features?}
\label{subsec:magnetographs}
We have identified two notable patterns in Dirichlet potential fields
extrapolated using SDO/HMI vector field measurements:
opposite-polarity halos around strong-field regions (middle-left panel
of Figure \ref{fig:bpot_comp}); and magnetic energies substantially
lower than Neumann potential fields.

Since these might arise from systematic effects in the observation and
inversion processes used to estimate the magnetic field by the SDO/HMI
Team, we also analyze magnetograms of two other active regions,
observed by the {\it Synoptic Optical Long-term Investigations of the
  Sun}/{\it Vector- SpectroMagnetograph} (SOLIS/VSM;
\citealp{Keller2003}) and by the {\it Solar Optical Telescope }/{\it
  SpectroPolarimeter} (SOT-SP) aboard {\em Hinode} \citep{Lites2013}.

\begin{figure}[ht]
\centerline{\includegraphics[width=6.5in]{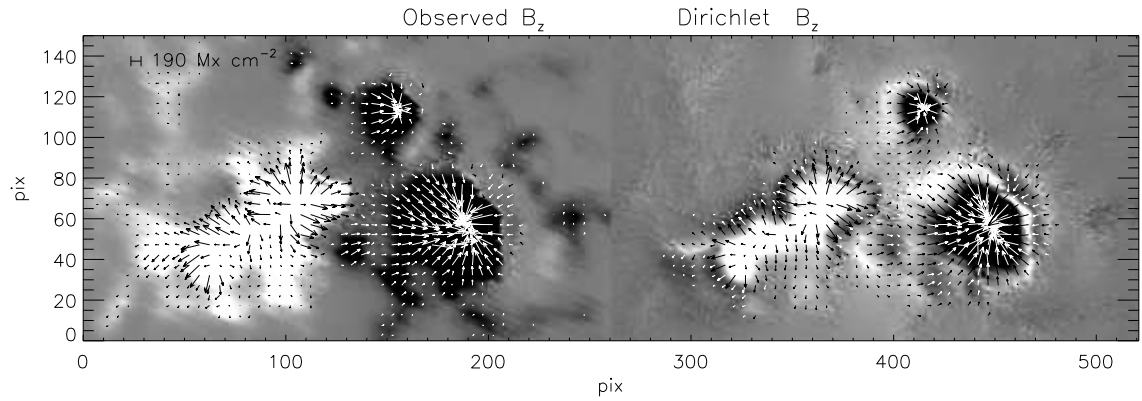}}
\caption{%
  Left: Grayscale shows vertical field $B_z$ in AR 11117
  observed by SOLIS/VSM on 27 October 2010 near 16:33 UT. Saturation
  is set to $\pm 125$ Mx cm$^{-2}$. Arrows (white where $B_z < 0$ and
  black where $B_z > 0$) show the observed horizontal field. Right:
  Grayscale shows vertical field $B_z$ from a potential extrapolation
  using the Dirichlet BC determined from the measured horizontal
  magnetic field. Arrows (white where $B_z < 0$ and black where $B_z >
  0$) show the potential horizontal field.
The $x$ and $y$ axes are in interpolated, reprojected HMI pixels,
which are square and 362 km on a side.
Note the opposite-polarity halos around strong-field regions in the
Dirichlet field, as in the SDO/HMI data.  }
\label{fig:solis}
\end{figure}

AR 11117 was observed by SOLIS/VSM on 27 October
2010. \cite{Tadesse2012} extrapolated the coronal magnetic field in
this region to determine the free magnetic energy content before a
C-class flare near 17:00 UT. Here, we compute the pre-flare potential
field energy from the VSM magnetogram recorded around 16:33 UT, which
is available from the National Solar Observatory's website. We used a
Lambert equal-area projection to remap the data onto a Cartesian
plane, with 1\arcsec{} pixels. In Figure \ref{fig:solis}, we show the
observed and Dirichlet potential fields. As with the SDO/HMI data in
Figure \ref{fig:bpot_comp}, opposite-polarity halos in $B_z$ are
present around strong-field regions in the Dirichlet solution.

\citet{Tadesse2012} adopt noise levels of 1 Mx cm$^{-2}$ and 50 Mx
cm$^{-2}$ for the line-of-sight and transverse field components,
respectively. Given the active region's location near N21W25 at the
time of these observations (about 32$^\circ$ from disk center),
transverse field measurements contribute about 25\% to the radial
field estimate. Hence, crude estimates of errors in the radial and
horizontal components would be about 12.5 Mx cm$^{-2}$ and 37.5 Mx
cm$^{-2}$, respectively, for field components in a plane through disk
center. (The horizontal error is probably higher, because one
component of the horizontal field does not include any line-of-sight
component. We have also neglected any discussion of ambiguity
resolution errors.) Assuming a 3:1 weighting in favor of the radial
field measurement, we find the Neumann, Dirichlet, and noise-weighted
fields have energies of $3.3 \times 10^{32}$ erg, $8.6 \times 10^{31}$
erg, and $2.4 \times 10^{32}$ erg, respectively. The Dirichlet energy
is smaller by a factor greater than 3, so the difference between the
Neumann and noise-weighted energies is nearly $10^{32}$ erg, about
half the free energy estimated \citep{Tadesse2012} from their
non-linear force-free-field (NLFFF) extrapolation. We remark that all
of our potenetial field energies are significantly smaller than the
potential-field energy estimate of \citet{Tadesse2012}, by roughly a
factor of 3. This might be due to the much larger field-of-view over
which they estimated the magnetic energy.

\begin{figure}[ht]
\centerline{\includegraphics[width=6.5in]{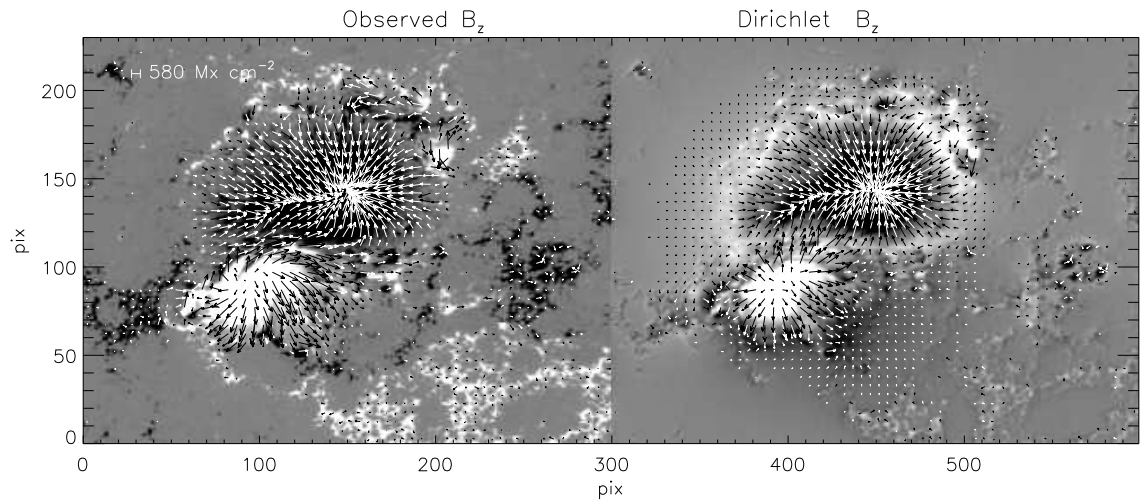}}
\caption{%
  Left: Grayscale shows vertical field $B_z$ in AR 10930
  observed by {\it Hinode}/SOT-SP on 12 December 2006 near 21:00
  UT. Saturation is set to $\pm 750$ G. Arrows (white where $B_z < 0$
  and black where $B_z > 0$) show the observed horizontal
  field. Right: Grayscale shows vertical field $B_z$ from a potential
  extrapolation using the Dirichlet BC determined from the measured
  horizontal magnetic field. Arrows (white where $B_z < 0$ and black
  where $B_z > 0$) show the potential horizontal field.
Pixels are 0.63\arcsec{} on a side.
Note the opposite-polarity halos around strong-field regions in the
Dirichlet field, as in the SDO/HMI and SOLIS/VSM data.  }
\label{fig:sp}
\end{figure}

\citet{Schrijver2008} analyzed NLFFF extrapolations of NOAA AR 10930
based upon a vector magnetogram derived from {\em Hinode}/SOT-SP
measurements near 21:00 UT on 12 December 2006. In the left panel of
Figure \ref{fig:sp}, we show the observed vector field ($B_z$ in
grayscale, horizontal field with vectors). (Pixels are 0.63\arcsec{}
on a side.) In this figure's right panel, we show the Dirichlet field
(again, $B_z$ in grayscale, horizontal field with vectors). As with
the SDO/HMI and SOLIS/VSM data, opposite-polarity halos are seen in
the vertical field around strong-field regions in the Dirichlet
solution. Perhaps not coincidentally, the large-scale, strong-field
polarities in the observed field (left panel) are surrounded by the
many small-scale concentrations of opposite-polarity flux. It is
possible that the photospheric magnetic field dynamically creates an
opposite polarity moat via a relaxation process toward a field that
more closely resembles the Dirichlet potential field. The observed
opposite-polarity moats are not, however, as intense as those in the
Dirichlet solution.

As with the SDO/HMI and SOLIS/VSM observations, the energy of the Neumann potential field for the {\it Hinode}/SOT-SP data, $\approx 1.6 \times 10^{33}$ erg, is also larger than the energy of the Dirichlet field, $\approx 8.1 \times 10^{32}$ erg, a difference of a factor near two. With a weighting of 10:1 in favor of the normal-field (Neumann) BC, the weighted field's potential energy is $\approx 1 \times 10^{32}$ erg less than the Neumann field's. This is roughly equivalent to the $\approx 10^{32}$ erg \citep{Emslie2012} estimated to have been released in an X-class flare and CME a few hours after this magnetogram was recorded. Again, we remark that different choices of BCs can significantly affect estimates of free energies in coronal fields.

\subsection{Hints about Photospheric Horizontal Currents}
\label{subsec:currents}
Beyond the quantitative differences between potential fields inferred using Neumann condition and hybrid BCs, a qualitative difference in the hybrid approach is its treatment of the observed $B_z$ as neither unrelated to the presence of currents nor ``gospel'' data that should be precisely matched when specifying the potential field.

In the hybrid approach, differences between the observed and potential $B_z$ both account for uncertainties in the measurements and indicate that horizontal currents are present at the photosphere. We can define the total field as the sum of potential and non-potential parts,
\be \vect{B}^{\rm tot} = \vect{B}^{\rm P} + \vect{B}^{\rm NP}
~. \ee
Then $B_z^{\rm NP}$ is related to horizontal current,
\bea
4 \pi J_x / c &=& \partial_y B_z - \partial_z B_y
= \partial_y B_z^{\rm NP} - \partial_z B_y^{\rm NP} \\
4 \pi J_y / c &=& \partial_z B_x - \partial_x B_z
= \partial_z B_x^{\rm NP} - \partial_x B_z^{\rm NP}
~. \eea
Magnetic field inversions that estimate the field at a single height cannot provide information about variation in magnetic field components along the LOS (or $z$) directions, so the structure of horizontal currents cannot be determined from these observations alone.

Nonetheless, patterns in the spatial structure of $B_z^{\rm NP}$ might reveal clues about horizontal photospheric currents. For instance, we have noted the opposite-polarity halos surrounding strong-field regions in the hybrid (or Dirichlet) potential fields. When differencing the observed and hybrid (or Dirichlet) potential normal magnetic fields in a sunspot --- as in, for instance, the bottom-left panels of Figures \ref{fig:bpot_comp} and \ref{fig:combo_comp} --- the halos around the positive-polarity spot at the upper right of the magnetogram produce an interesting structure for $B_z^{\rm NP}$: a core field with opposite polarity to that of the spot, surrounded by a moat of the same polarity as the spot. This structure would be produced by a horizontal electric current flowing clockwise (CW) around the spot's periphery. This current would flow in the opposite sense of the counterclockwise (CCW) sheath current expected to flow around an isolated, positive flux tube surrounded by field-free plasma ({\it e.g.}, \citealp{Spruit1981}). \citet{Bommier2011} have also reported similar current structures in another active region, deduced using a different approach: multiple-height vector magnetic field measurements.

It should be noted that our interpretation of these non-potential field patterns as manifestations of {\em horizontal} currents implies that they are distinct from non-potential field patterns associated with {\em vertical} currents observed to cross the photosphere ({\it e.g.}, sheared, strong-field polarity inversion lines; \citealp{Georgoulis2012}). Hence, the relationship between the horizontal currents that we describe and vertical currents, whether outgoing/ ``direct'' or inflowing/``return'' currents ({\it e.g.}, \citealp{Torok2014}), is unclear.

\begin{figure}[ht]
\centerline{\includegraphics[width=5.5in]{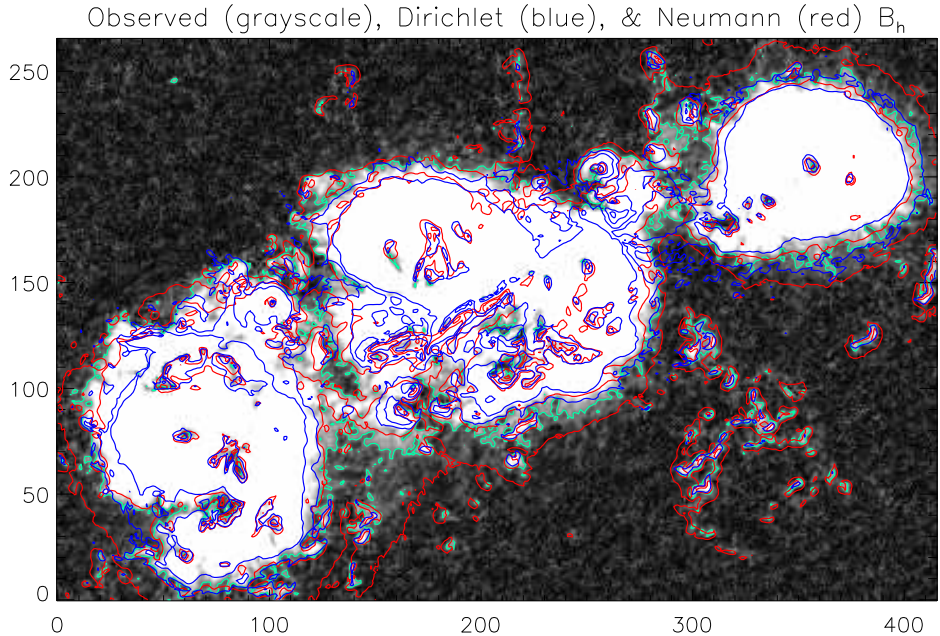}}
\caption{%
  The grayscale shows the smoothed horizontal field strength $|\vect
  B_h|$ from the SDO/HMI observation in Figure \ref{fig:bpot_comp}.
The $x$ and $y$ axes are in interpolated, reprojected HMI pixels,
which are square and 362 km on a side.
The aqua contour line shows the 200 Mx cm$^{-2}$ level curve of the
observed $|\vect{B}_h|$. The red and blue contour lines show 200 and
400 Mx cm$^{-2}$ level curves of $|\vect{B}_h|$ from the Neumann and
Dirichlet potential fields, respectively. Saturation of the grayscale
is set to 400 Mx cm$^{-2}$. Note the presence of significant
horizontal fields in the Neumann solution outside of the observed
horizontal field ({\it i.e.}, outside the aqua line).  }
\label{fig:bh_comp}
\end{figure}

It is also possible that the halos arise from the unphysical nature of
the potential field assumption: the actual active-region field is
confined by surrounding, quiet-sun regions, but potential fields are
by definition force-free, so are space-filling. In fact, Neumann
potential fields typically exhibit this effect: Neumann solutions tend
to have strong, spatially coherent horizontal magnetic fields outside
of intense concentrations of vertical field. This contrasts to the
observed horizontal fields, which tend to be much more strongly
confined to areas with strong vertical flux: the observed horizontal
fields in Figure \ref{fig:bh_comp}, where $|\vect{B}_h|$ from the
SDO/HMI dataset is shown in grayscale, closely match areas where
strong vertical fields are present in the upper-left panel of Figure
\ref{fig:bpot_comp}. A moat of extended horizontal fields from the
Neumann solution can also be seen in Figure \ref{fig:bh_comp}, where
the 200 Mx cm$^{-2}$ contour line of the Neumann field's $|\vect{B}_h|$
(the outermost red contour) lies well outside the same-strength
contours of both $|\vect{B}_h|$ observed by SDO/HMI (aqua contour) and
the Dirichlet field $|\vect{B}_h|$ (blue contour). These ``excess''
horizontal fields in the Neumann solution indicate that, relative to
the potential-field model, the actual magnetic field is ``too
vertical.'' The Dirichlet field's dearth of vertical flux (compared to
the observed field) accords with this characterization: relative to
the vertical field in the Dirichlet model, which is derived from the
observed horizontal field, the actual field's area-integrated $|B_z|$
is higher, {\it i.e.}, the real field tends to be more vertical.

We suspect that opposite-polarity the halos present in the Dirichlet
field's (modeled) $B_z$ likely arise from the rapid fall-off observed
in horizontal field strengths, which tend to produce halos in $\divgh
\vect{B}_h$ like those discernible in the middle-right panel of Figure
\ref{fig:bpot_comp}. Simply put: halos in the observed $\divgh \vect
B_h$ produce halos in the Dirichlet model's $B_z$.

Despite the presence of opposite-polarity halos elsewhere in Figures
\ref{fig:bpot_comp} -- \ref{fig:combo_comp} and in Figures
\ref{fig:solis} -- \ref{fig:sp}, similar structures in $B_z^{\rm NP}$
are not ubiquitous in other areas of the magnetograms we have analyzed
here. Studying Dirichlet / hybrid potential magnetic fields in
additional active regions, as well as additional multi-height vector
magnetic field measurements, would be useful to better understand the
structure of electric currents within active region fields.

\section{Summary and Discussion}
\label{sec:conc}
We have presented methods to compute potential (current-free) magnetic
fields in the photosphere and corona that incorporate measurements of
the horizontal photospheric magnetic field. The horizontal
photospheric field determines a 2D Dirichlet boundary condition at the
photosphere to solve Laplace's equation in 3D. In addition to matching
just the horizontal field, we also described methods that attempt to
match both vertical and horizontal components of the field, in a
statistical sense.

Historically, potential magnetic fields have been computed using estimates of just the vertical photospheric magnetic field, which determines a Neumann boundary condition there for Laplace's equation in three dimensions. This was perhaps because the line-of-sight magnetic field can be measured more easily and accurately than the component of the magnetic field transverse to the LOS, and near disk center the LOS field approximates the radial field. In fact, the difficulty of measuring the full magnetic vector at the photosphere meant that such observations were relatively rare prior to the advent of SOLIS/VSM, {\em Hinode}/SOT-SP, and SDO/HMI. Given the expected presence of horizontal electric currents in active regions at the photosphere (indeed, throughout the photosphere-to-corona volume), however, there is no physical basis for assuming that a potential field's vertical component should precisely match that of the observed field. Hence, the customary use of Neumann boundary conditions for potential field extrapolations likely arose from observational capabilities, not physical considerations.

We showed that a hybrid field, formed by a linear combination of separate solutions to Laplace's equation using the Neumann and Dirichlet boundary conditions, matches the observed vector magnetic field more closely than either the Neumann or Dirichlet solutions alone. If the aim of determining a potential field is to find the current-free coronal magnetic field {\em most consistent with the observed photospheric field,} then evidently information about the horizontal field should also be included when deriving the potential field. Differences between the observed field and the hybrid field that most closely matches it can arise from noise in the magnetic field measurements, but are also expected in the presence of electric currents.

We found two systematic features of Dirichlet and hybrid fields. First, we noted that opposite-polarity halos of vertical magnetic field surrounding strong-field concentrations of vertical flux are typical in the magnetograms we studied. These are not seen in the observed vertical field. Second, we found that the magnetic energy of the Dirichlet and hybrid potential fields tends to be significantly lower than that of the Neumann potential fields. The differences in magnetic energies of these potential fields --- 10$^{32}$ erg or more --- are commensurate with estimates of the energy released in solar eruptive events ({\it e.g.}, \citealp{Emslie2012}). (As noted above, the assumption of periodicity implicit in our use of DFTs to compute potential fields might yield energies that are artificially large.) We found that Dirichlet or hybrid potential fields derived from SDO/HMI, SOLIS/VSM, and {\em Hinode}/SOT-SP vector magnetograms all exhibited both properties.

Our results have two notable implications for estimates of free magnetic energy in non-potential coronal magnetic field models. First, from a physical point of view, the free magnetic energy necessarily depends upon the assumed minimum-energy field, and hybrid potential fields both match observations more closely than Neumann potential fields and have lower energy. This suggests that estimates of coronal free energy made using the Neumann field as minimum-energy field might understate the true free energy significantly.

Second, from a practical standpoint, hybrid potential fields probably provide better initial conditions than Neumann fields for some methods to calculate model coronal fields. For instance, the initial coronal field in the optimization method ({\it e.g.}, \citealp{Wheatland2000a, Wiegelmann2008}) of deriving NLFFFs is typically extrapolated from a photospheric Neumann condition, but the relaxation is initiated by replacing the potential photospheric boundary field with the observed field (or a ``preprocessed'' version of the observed field). A large discrepancy between the observed boundary field and the model field's first interior layer is then present. The optimization algorithm then attempts to remove any Lorentz forces or magnetic divergences due to this discrepancy by iteratively altering the model field above the photosphere. Because hybrid-potential photospheric fields more closely match the observed photospheric field, the optimization algorithm should converge more quickly starting from a hybrid potential solution, since discrepancies between the initial and observed $\divgh \vect{B}_h$ should be smaller. This can be understood by considering a Helmholtz decomposition of $\vect{B}_h$, which expresses $\vect{B}_h$ in terms of a scalar potential and a stream function: the scalar potential is derived from $\divgh \vect{B}_h$, and can be used to determine the irrotational part of $\vect{B}_h$; the stream function is derived from the solenoidal component of $\vect{B}_h$, from $\hatz \cdot (\curlh \vect{B}_h)$, and can be used to determine the remaining part of $\vect{B}_h$ (which is directly related to the vertical photospheric electric current). (This decomposition underlies a poloidal-toroidal decomposition [PTD] of the full, three-component vector, $\vect{B}$; \citet{Fisher2010} and \citep{Kazachenko2014} discuss PTD at length, and include detailed descriptions of methods to determine the scalar potential and stream function.) Essentially, NLFFF methods should be focused on matching the observed vertical currents, related to $\hatz \cdot (\curlh \vect{B}_h)$, not discrepancies between a Neumann-derived $\divgh \vect{B}_h$ versus the observed $\divgh \vect{B}_h$. In fact, it is likely that discrepancies in $\divgh \vect{B}_h$ introduce significant problems in NLFFF extrapolations: \citet{Valori2013} report that errors in satisfying $\nabla \cdot \vect{B} = 0$ in extrapolated fields can compromise magnetic energy estimates, and starting from a potential extrapolation that is more consistent with the observed $\divgh \vect{B}_h$ should minimize introduction of spurious divergences. For similar reasons, we expect that using hybrid-field initial states in magnetofrictional methods ({\it e.g.}, \citealp{Valori2010}) could also result in closer agreement between observed and extrapolated fields.

As another consequence of using a hybrid potential field, some treatments of magnetic helicity, such as those developed by \citet{Berger1984} and \cite{Finn1985}, are not applicable with hybrid fields. These treatments sought to define a gauge-invariant helicity for a given magnetic field, $\vect{B}$, by defining that field's helicity ``relative'' to the ``reference'' potential magnetic field satisfying the corresponding Neumann boundary condition. For these definitions of helicity, precise agreement between the actual and potential fields' normal components is required for gauge invariance. We note, however, that other formalisms that yield well-defined values for magnetic helicity have been developed ({\it e.g.}, \citealp{Hornig2006, Low2015}) that do not require any reference field.

For simplicity, we used DFTs to compute potential fields discussed above. In the Appendices, we present methods for the more general Cartesian cases of non-periodic Neumann and Dirichlet potential fields, using finite-element sources. For the global spherical case, \citet{Nemenman1999} present Green's functions for the 3D Neumann and Dirichlet boundary conditions (without a source-surface boundary condition), which can be used to match observations with arbitrary spatial resolution. To derive the Dirichlet condition from observations, the 2D Poisson problem in spherical coordinates can be solved using FISHPACK, a suite of Fortran codes developed at NCAR by \citet{Schwarztrauber1975}. An IDL wrapper for FISHPACK has been developed as part of the CGEM project (http://cgem.stanford.edu/), and should be publicly released soon. We note that, in principle, spherical harmonic transforms could be used for either of these 3D or 2D problems, but the numerical expense of computing very-high-order expansions to match high-resolution observations can be prohibitive.

\acknowledgements
We thank the referee for providing valuable comments that improved this paper, especially pointing out the incompatibility of non-Neumann potential fields with relative helicity. BTW thanks T. Sakurai for insight gained from a discussion of this work at the Hinode-7 meeting. We acknowledge funding from the NSF's National Space Weather Program under award AGS-1024862, the NASA Living-With-a-Star TR\&T Program (grant NNX11AQ56G), the NASA Heliophysics Theory Program (grant NNX11AJ65G), and the Coronal Global Evolutionary Model (CGEM) project, funded by award NSF AGS 1321474. NASA/SDO and the HMI instrument were joint efforts by many teams and individuals, whose efforts to produce the HMI magnetograms that we analyzed here are greatly appreciated. SOLIS/VSM vector magnetograms are produced cooperatively by NSF/NSO and NASA/LWS. The National Solar Observatory (NSO) is operated by the Association of Universities for Research in Astronomy, Inc., under cooperative agreement with the National Science Foundation. {\em Hinode} is a Japanese mission developed and launched by ISAS/JAXA, collaborating with NAOJ as a domestic partner, and NASA and STFC (UK) as international partners. Scientific operation of the {\em Hinode} mission is conducted by the {\em Hinode} science team organized at ISAS/JAXA. This team mainly consists of scientists from institutes in the partner countries. Support for the post-launch operation is provided by JAXA and NAOJ (Japan), STFC (UK), NASA (USA), ESA, and NSC (Norway). The authors are grateful to the U.S. taxpayers for providing the funds necessary to perform this work.
%\end{acks}

\appendix

\section{Dirichlet Solution for a Tile}
\label{sec:dirich_tile}
For some applications, the periodicity inherent in using DFTs to calculate 3D potential fields from the 2D Dirichlet BC is problematic. Here, we outline an approach to determine a non-periodic, 3D potential function in Cartesian geometry. In a domain that is finite in $x$ and $y$, one can calculate the 2D Dirichlet BC, $\chi(x,y,0)$, for the 3D potential function $\chi(x,y,z)$ directly from the boundary values of $\divgh \vect{B}_h(x_i,y_i) \vert_{z=0}$, without solving the 2D Poisson's equation. The approach is similar to using a Green's function: superposition is employed with the fundamental solution to the 2D Laplace equation,
\be \chi_{\rm D, pt}(x,y) = \frac{-D_{\rm pt}}{2 \pi}
\ln ( \sqrt{(x - x')^2 + (y - y')^2} )
~, \label{eqn:fundamental} \ee
which gives the potential at $(x,y)$ due to a point divergence of strength $D_{\rm pt}$ at $(x',y')$. The divergence of the 2D vector field computed from the horizontal gradient of this potential vanishes for all points other than $(x',y')$. With the horizontal divergence expressed as a 2D function $D(x,y) = \divgh \vect{B}_h(x_i,y_i) \vert_{z=0}$, the potential at $z=0$, $\chi_{\rm D}(x,y,0)$ is found by summing the contributions from all sources over the 2D domain. (Unlike the situation with a Green's function, an integral over the fundamental solution does not converge on an infinite domain.)

Once the potential function is specified on $z=0$, the potential at an arbitrary point in the domain $(x,y,z)$ can be computed {\it via}
\be \chi(x,y,z) = \frac{z}{2\pi} \int {\rm d}A' \frac{\chi(x', y', 0)}
{[(x - x')^2 + (y - y')^2 + z^2]^{3/2}}
~. \label{eqn:dirichlet_3d} \ee
It should be noted that the potential at given height $H$ above a point $(x_0, y_0)$ on the surface will be significantly influenced by values of the potential on the surface {\em at least} a horizontal distance $H$ away from $(x_0, y_0)$. Therefore, to extrapolate the field above a rectangular region of the photosphere $N_x \times N_y$ pixels in extent to a height commensurate with the area, $H \approx \sqrt{N_x \times N_y}$, the potential $\chi_{\rm D}(x,y,0)$ must be known over a substantially larger area of the $z=0$ plane --- of order $3 N_x \times 3 N_y$ or more. In contrast, when extrapolating from a Neumann BC in a region, the normal field exterior to that region is often assumed to vanish, meaning the Neumann extrapolation does not require input from so large an area. Hence, the Dirichlet field's dependence upon the potential at the surface, which in general extends far outside the Neumann field's sources, is a disadvantage in practical terms. One possible work-around is to: determine $\chi_{\rm D}(x,y,0)$ over an area only slightly larger than the $N_x \times N_y$ region; extrapolate upward just enough zones to determine $\partial_z \chi(x,y,0)$ from a one-sided, finite-difference stencil for the vertical derivative; and then use {\em this} normal derivative ({\em not} the observed normal field) as a Neumann boundary condition to extrapolate the potential field.

One practical problem in computing $\chi_{\rm D}(x,y,0)$, however, is that the fundamental solution is singular at a given source's location. For numerical determination of Dirichlet potential fields (for instance, to integrate field lines), this is a major shortcoming. To overcome this deficiency, one can treat the horizontal divergence computed for the pixel $D_{ij}$ at $(x_i,y_j)$ as that due to an extended source with spatially uniform areal density $\bar D = D_{ij}/(\Delta s)^2$ over a square ``tile'' of finite extent $\Delta s$ in $x$ and $y$, corresponding to the pixel area. The resulting expression for the potential due to this finite source is not singular at the pixel center, or elsewhere (for a finite domain in $x$ and $y$).

We now determine the finite-element potential, assuming the pixel is centered at the origin. It suffices to compute the potential at two points: (i) at the center of the tile; and (ii) at an arbitrary point outside the tile, which we will use to compute the potential at neighboring pixels. For points far from the tile, relative to $\Delta s$, we can approximate the potential as that from a point source.

\subsection{Solution at Tile Center}
\label{subsec:dtile_in}
We first want to compute the potential corresponding to the point at the center of the tile,
\be \chi_{\rm D}(0,0)
=
\int_{-\Delta s/2}^{\Delta s/2} {\rm d}x' \int_{-\Delta s/2}^{\Delta s/2} {\rm d}y'
\frac{-\bar D}{2 \pi} \ln( \sqrt{x'^2 + y'^2})
~, \ee
where the argument of $\chi_{\rm D}$ refers to distance with respect to the center of the source pixel, the origin for $x'$ and $y'$. Then
\bea \chi_{\rm D}(0,0)
&=& \frac{-\bar D}{4 \pi} \int_{-\Delta s/2}^{\Delta s/2} {\rm d}x' \int_{-\Delta s/2}^{\Delta s/2} {\rm d}y'
\ln (x'^2 + y'^2) \\
&=& \frac{-\bar D}{4 \pi} \int_{-\Delta s/2}^{\Delta s/2} {\rm d}x'
\left . \left [y' \ln (x'^2 + y'^2) - 2y' + 2 x' \tan^{-1}\frac{y'}{x'}
\right ] \right \vert_{-\Delta s/2}^{\Delta s/2} \\
&=& \frac{-\bar D}{4 \pi} \int_{-\Delta s/2}^{\Delta s/2} {\rm d}x'
\left [\Delta s \ln (x'^2 + (\Delta s)^2/4) - 2 \Delta s + 4 x' \tan^{-1}\frac{\Delta s}{2x'}
\right ] \\
&=& \frac{-\bar D}{4 \pi} \left \{ \int_{-\Delta s/2}^{\Delta s/2} {\rm d}x'
\left [\Delta s \ln (x'^2 + (\Delta s)^2/4) - 2 \Delta s \right ] \right . \nn \\
&& \left . + 4 \int_{-\Delta s/2}^{0} {\rm d}x' x' \tan^{-1}\frac{\Delta s}{2x'}
+ 4 \int_{0}^{\Delta s/2}  {\rm d}x' x' \tan^{-1}\frac{\Delta s}{2x'} \right \} \\
&=& \frac{-\bar D}{4 \pi} \left \{ \int_{-\Delta s/2}^{\Delta s/2} {\rm d}x'
\left [\Delta s \ln (x'^2 + (\Delta s)^2/4) - 2 \Delta s \right ]  +
8 \int_{0}^{\Delta s/2} {\rm d}x' x' \cot^{-1}\frac{x'}{\Delta s/2} \right \} \\
&=& \frac{-\bar D}{4 \pi} \left \{
\left . \left [\Delta s x' \ln (x'^2 + (\Delta s)^2/4) - 2 \Delta s x'
+ (\Delta s)^2 \tan^{-1}\frac{x'}{\Delta s/2}
- 2 \Delta s x' \right ] \right \vert_{-\Delta s/2}^{\Delta s/2}  \right . \nn \\
&& + \left . \left .
8 \left[\frac{1}{2}(x'^2 + (\Delta s)^2/4)\cot^{-1}\frac{x'}{\Delta s/2}
+ \frac{\Delta s x'}{4} \right ]
\right \vert_{0}^{\Delta s/2}  \right \} \\
&=& \frac{-\bar D}{4 \pi} \left \{
(\Delta s)^2 \ln ((\Delta s)^2/2) - 4 (\Delta s)^2
+ (\Delta s)^2 \pi [1/2 + 1/2 - 1/2]
+ (\Delta s)^2 \right \} \\
&=& \frac{-\bar D (\Delta s)^2 }{4 \pi} \left \{
2 \ln (\Delta s) - \ln (2)
- 3 + \pi/2 \right \} \\
&=& \frac{-D_{ij}}{4 \pi} \left \{
2 \ln (\Delta s) - \ln (2)
- 3 + \pi/2 \right \}
= \frac{-D_{ij}}{4 \pi} \left \{
2 \ln (\Delta s) - 2.122 \right \}
~. \eea

\subsection{Solution Outside of Tile}
\label{subsec:dtile_out}
For the potential exterior to the tile, but near it, we now want to compute
\be \chi_{\rm D}(x,y)
=
\int_{-\Delta s/2}^{\Delta s/2} {\rm d}x' \int_{-\Delta s/2}^{\Delta s/2} {\rm d}y'
\frac{-\bar D}{2 \pi} \ln( \sqrt{(x - x')^2 + (y - y')^2})
~, \label{eqn:outside} \ee
where, as above, the arguments of $\chi_{\rm D}$ refers to distance with respect to the center of the source pixel, which we take as the origin.

We first change variables, $x'' = x - x'$ and $y'' = y - y'$, and define
\bea x_\pm &=& x \pm \Delta s/2 \label{eqn:xpm} \\
y_\pm &=& y \pm \Delta s/2 ~. \label{eqn:ypm} \eea
Then the upper and lower limits of integration $\pm \Delta/2$ for $x'$ become $x_\mp$ for $x''$, and similarly for $y'$ and $y''$. We then have
\bea \chi_{\rm D}(x,y)
&=& \frac{-\bar D}{4 \pi} \int_{x_-}^{x_+} {\rm d}x'' \int_{y_-}^{y_+} {\rm d}y''
\ln (x''^2 + y''^2) \\
&=& \frac{-\bar D}{4 \pi} \int_{x_-}^{x_+} {\rm d}x''
\left . \left ( y'' \ln (x''^2 + y''^2) - 2y'' + 2 x'' \tan^{-1}\frac{y''}{x''}
\right ) \right \vert_{y_-}^{y_+} \\
&=& \frac{-\bar D}{4 \pi} \int_{x_-}^{x_+} {\rm d}x''
\left ( \left [ y_+ \ln (x''^2 + y_+^2) - 2y_+
+ 2 x'' \tan^{-1}(\frac{y_+}{x''}) \right ] \right . \nn \\
&& \phantom{\frac{-\bar D}{4 \pi} \int_{x_-}^{x_+} {\rm d}x''} \left .
\!\! - \left [ y_- \ln (x''^2 + y_-^2) - 2y_-
+ 2 x'' \tan^{-1}(\frac{y_-}{x''}) \right ] \right ) \\
&=& \frac{-\bar D}{4 \pi} \int_{x_-}^{x_+} {\rm d}x'' \left ( - 2 \Delta s
+ \left [ y_+ \ln (x''^2 + y_+^2) + 2 x'' \cot^{-1}\frac{x''}{y_+}
 \right ]  \right . \nn \\
&& \phantom{\frac{-\bar D}{4 \pi} \int_{x_-}^{x_+} {\rm d}x'' \left ( - 2 \Delta s \right .}
\left .
- \,\, \left [ y_- \ln (x''^2 + y_-^2) + 2 x'' \cot^{-1}\frac{x''}{y_-} \right ]
\right ) ,
\eea

\bea
\lefteqn{\chi_{\rm D}(x,y) = \frac{-\bar D}{4 \pi}
\left ( \vphantom{\tan^{-1}\frac{x''}{y_+}} - 2 \Delta s x'' \right .  } \nn \\
&+&
\left [ y_+ \left ( x'' \ln (x''^2 + y_+^2) - 2 x'' + 2 y_+ \tan^{-1}\frac{x''}{y_+}
\right )
+ (x''^2 + y_+^2) \cot^{-1}\frac{x''}{y_+} + x''y_+
 \right ]  \nn \\
&-& \left . \left .
\left [ y_- \left ( x'' \ln (x''^2 + y_-^2) - 2 x'' + 2 y_- \tan^{-1}\frac{x''}{y_-}
\right )
+ (x''^2 + y_-^2) \cot^{-1}\frac{x''}{y_-} + x''y_-
\right ]
\right ) \right \vert_{x_-}^{x_+} , \eea

\bea
\lefteqn{\chi_{\rm D}(x,y) = \frac{-\bar D}{4 \pi}
\left ( \vphantom{\tan^{-1}\frac{x''}{y_+}} - 2 \Delta s^2 \right . } \nn \\
&+&
\left [ y_+ \left ( x_+ \ln (x_+^2 + y_+^2) - 2 x_+ + 2 y_+ \tan^{-1}\frac{x_+}{y_+}
\right )
+ (x_+^2 + y_+^2) \cot^{-1}\frac{x_+}{y_+} + x_+y_+
 \right ]  \nn \\
&-&
\left [ y_- \left ( x_+ \ln (x_+^2 + y_-^2) - 2 x_+ + 2 y_- \tan^{-1}\frac{x_+}{y_-}
\right )
+ (x_+^2 + y_-^2) \cot^{-1}\frac{x_+}{y_-} + x_+y_-
\right ]  \nn \\
&-&
\left [ y_+ \left ( x_- \ln (x_-^2 + y_+^2) - 2 x_- + 2 y_+ \tan^{-1}\frac{x_-}{y_+}
\right )
+ (x_-^2 + y_+^2) \cot^{-1}\frac{x_-}{y_+} + x_-y_+
 \right ]  \nn \\
&+& \left .
\left [ y_- \left ( x_- \ln (x_-^2 + y_-^2) - 2 x_- + 2 y_- \tan^{-1}\frac{x_-}{y_-}
\right )
+ (x_-^2 + y_-^2) \cot^{-1}\frac{x_-}{y_-} + x_-y_-
\right ]
\right ) \
~. \eea

This expression can be recast in terms of the measured horizontal divergence $D_{ij}$ at $(x_i,y_j)$ and pixel indices $(i,j)$ as
\bea
\lefteqn{\chi_{\rm D}(x,y) = \frac{-D_{ij}}{4 \pi}
\left ( \vphantom{\tan^{-1}\frac{i''}{j_+}} 2 \ln(\Delta s) - 2  \right . } \nn \\
&+&
\left [ j_+ \left ( i_+ \ln (i_+^2 + j_+^2) - 2 i_+ + 2 j_+ \tan^{-1}\frac{i_+}{j_+}
\right )
+ (i_+^2 + j_+^2) \cot^{-1}\frac{i_+}{j_+} + i_+j_+
 \right ]  \nn \\
&-&
\left [ j_- \left ( i_+ \ln (i_+^2 + j_-^2) - 2 i_+ + 2 j_- \tan^{-1}\frac{i_+}{j_-}
\right )
+ (i_+^2 + j_-^2) \cot^{-1}\frac{i_+}{j_-} + i_+j_-
\right ]  \nn \\
&-&
\left [ j_+ \left ( i_- \ln (i_-^2 + j_+^2) - 2 i_- + 2 j_+ \tan^{-1}\frac{i_-}{j_+}
\right )
+ (i_-^2 + j_+^2) \cot^{-1}\frac{i_-}{j_+} + i_-j_+
 \right ]  \nn \\
&+& \left .
\left [ j_- \left ( i_- \ln (i_-^2 + j_-^2) - 2 i_- + 2 j_- \tan^{-1}\frac{i_-}{j_-}
\right )
+ (i_-^2 + j_-^2) \cot^{-1}\frac{i_-}{j_-} + i_-j_-
\right ]
\right ) \
~, \eea
where $i_\pm$ is the numerical value of the pixel index $i$ plus or minus 1/2 ({\it i.e.}, {\em not} the address of a data value midway between two pixels), and similarly for $j_\pm$. The pixel length scale $\Delta s$ enters in just one term.

\begin{figure}[ht]
\centerline{\includegraphics[width=4.25in]{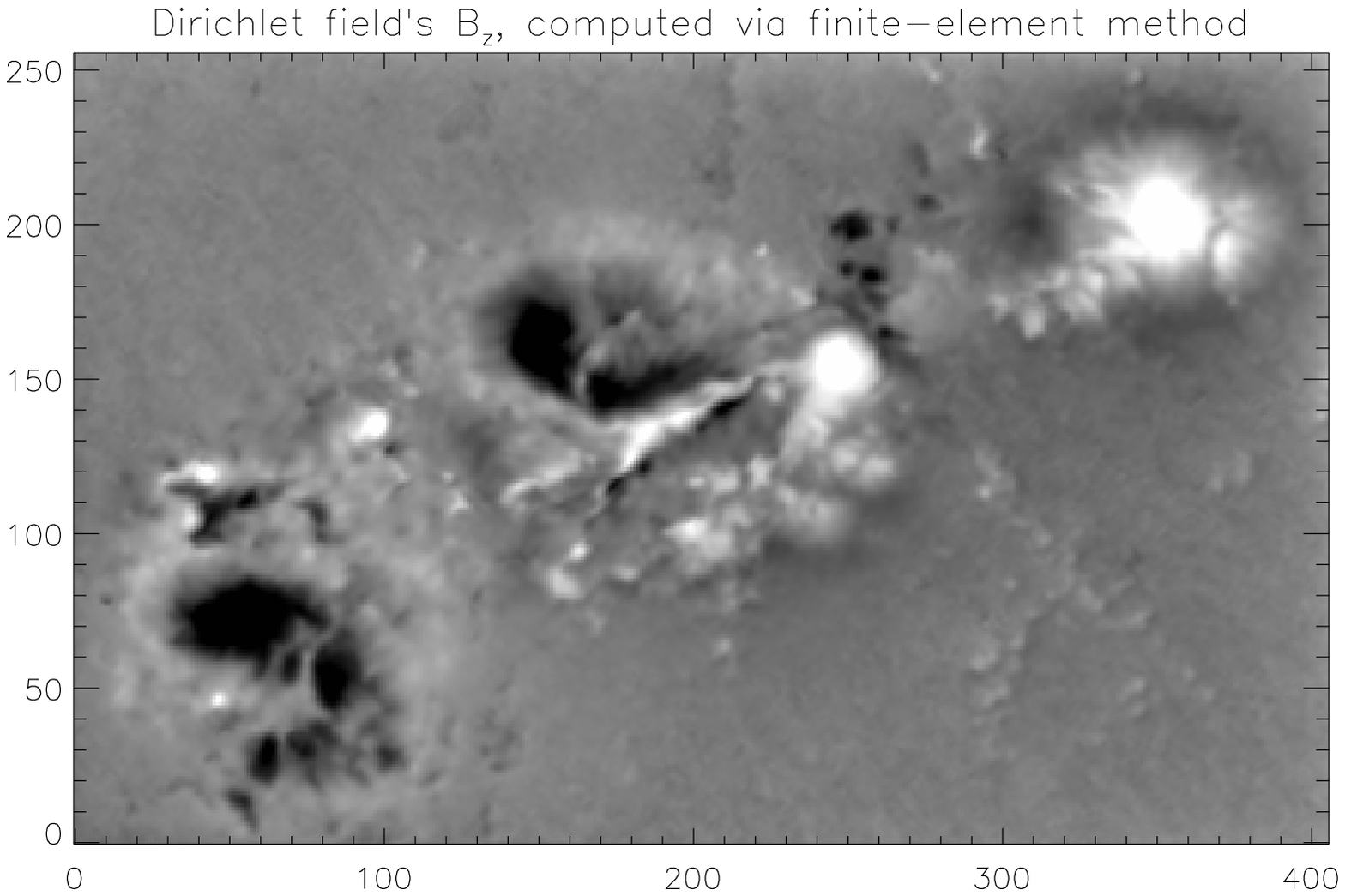}}
\caption{%
  Grayscale map of $B_z$ in AR 11158 from the Dirichlet BC
  computed using the finite-element (``tile'') approach derived in
  Appendix \ref{sec:dirich_tile}.
  The $x$ and $y$ axes are in interpolated, reprojected HMI pixels,
which are square and 362 km on a side.
Saturation is set to $\pm 1000$ Mx cm$^{-2}$. The opposite-polarity
halos present around strong-field regions in the Dirichlet magnetic
field computed with DFTs (see middle-left panel of figure
\ref{fig:bpot_comp}) are also visible here.  }
\label{fig:bz_fe}
\end{figure}

We give an example of $B_z$ computed with the finite-element approach in Figure \ref{fig:bz_fe}, which shows the Dirichlet field's $B_z$ in AR 11158. The method is numerically expensive (compute time scales as $N^4$ for each 2D slice in $z$), so we only computed the field on a 426 $\times$ 276 cropped region of the full 621 pixel $\times$ 610 array. We exclude a 10-pixel-wide strip around the cropped array's perimeter from this image, since the solution in much of this strip was strongly influenced by edge effects (owing to the cropping). This solution also exhibits the opposite-polarity halos present around strong-field regions that were visible in the Dirichlet field computed with DFTs (see middle-left panel of Figure \ref{fig:bpot_comp}).

\section{Neumann Solution for a Tile}
\label{sec:neum_tile}
Using hybrid Neumann-Dirichlet potential fields requires computing the Neumann and Dirichlet potentials with consistent approaches. Hence, in addition to the method presented above for computing the Dirichlet potential from a finite source --- a uniform-density tile --- we also describe a method to compute the potential arising from a tile of vertical flux, the analogous source for the Neumann case. As above, we use superposition to compute the potential from tile, building upon the Green's function for the 3D Laplace equation,
\be \chi_{\rm N, pt}(x,y,z) =
\frac{1}{2 \pi}
\frac{\Phi_{\rm pt}}{\sqrt{(x - x')^2 + (y - y')^2 + z^2}}
~, \label{eqn:neum_greens} \ee
which gives the potential at $(x,y,z)$ due to a point source of flux $\Phi_{\rm pt}$ at $(x',y',0)$. As with the Dirichlet case, this potential is singular at a point source's location, which can be problematic for numerical determination of Neumann potential fields. We again treat the flux $\Phi_{ij}$ in a pixel at $(x_i,y_j)$ as that due to an extended source with spatially uniform areal density $\bar B_z = \Phi_{ij}/(\Delta s)^2$ over a square tile of finite extent $\Delta s$ in $x$ and $y$, corresponding to the pixel area. As with the Dirichlet case, the resulting expression for the potential due to this finite source is not singular at the pixel center.

We now determine the finite-element potential, assuming the pixel is centered at the origin. It suffices to compute the potential at two points: (i) at the center of the tile; and (ii) at an arbitrary point outside the tile, which we will use to compute the potential at neighboring pixels. For points far from the tile, relative to $\Delta s$, we can approximate the potential as that from a point source. The total potential arising from an arbitrary distribution of fluxes over a set of pixels can then be computed by summing the potentials from all tiles.

\subsection{Solution at Tile Center}
\label{subsec:ntile_in}
We first want to compute the potential corresponding to points at the center of the tile,
\be \chi_{\rm N}(0,0)
= \frac{\bar B_z}{2 \pi}
\int_{-\Delta s/2}^{\Delta s/2} {\rm d}x' \int_{-\Delta s/2}^{\Delta s/2} {\rm d}y'
\frac{1}{\sqrt{x'^2 + y'^2}}
~.\ee
Defining $r = \sqrt{x'^2 + y'^2}$ and $\theta = \tan^{-1}(y'/x')$ gives
\be \chi_{\rm N}(0,0)
= \frac{\bar B_z}{2 \pi} \left (
\int_{0}^{2\pi} {\rm d}\theta \int_0^{\Delta s/2} {\rm d}r +
4 \int_{\Delta s/2}^{\sqrt 2 \Delta s/2} {\rm d}r
\int_{\theta_-(r)}^{\theta_+(r)} {\rm d}\theta \right)
~, \ee
where $\theta_- = \cos^{-1}(\Delta s/2r)$ and $\theta_+ = \sin^{-1}(\Delta s/2r)$. Then
\be \chi_{\rm N}(0,0)
= \frac{-\bar B_z}{2 \pi} \left ( \pi \Delta s +
4 \int_{\Delta s/2}^{\sqrt 2 \Delta s/2} {\rm d}r
[ \sin^{-1}(\Delta s/2r) - \cos^{-1}(\Delta s/2r)] \right)
~. \ee
Defining $u = (\Delta s/2)/r$, so ${\rm d}r = -(\Delta s/2) u^{-2} {\rm d}u$,
\bea \chi_{\rm N}(0,0)
&=& \frac{\bar B_z}{2 \pi} \left ( \pi \Delta s +
2 \Delta s  \int^{1}_{1/\sqrt 2} {\rm d}u u^{-2}
[ \sin^{-1}(u) - \cos^{-1}(u)] \right)  \\
&=& \frac{\bar B_z}{2 \pi} \left ( \pi \Delta s +
2 \Delta s  \left
[-\frac{1}{u} \sin^{-1}(u) - \ln \left ( \frac{1 + \sqrt{1 - u^2}}{u} \right )
\right . \right .
\nn \\
&& \left .  \left . \left .
+\frac{1}{u} \cos^{-1}(u) - \ln \left ( \frac{1 + \sqrt{1 - u^2}}{u} \right )
\right ] \right \vert^{1}_{1/\sqrt 2} \right)  \\
&=& \frac{\bar B_z}{2 \pi} ( \pi \Delta s +
2 \Delta s  [-\pi/2 + 2 \ln ( 1 + \sqrt{2} ) ] )  \\
&=& \frac{2 \bar B_z \Delta s }{ \pi} \ln ( 1 + \sqrt{2} )
~. \eea
This result implies that, to compute the potential at the point corresponding to the center of a pixel, assuming a uniform distribution of flux over that pixel, the denominator in Equation (\ref{eqn:neum_greens}) --- which is zero at that point --- should be replaced with $\Delta s/(4 \ln[\sqrt 2 + 1])$.

\subsection{Solution Outside of Tile}
\label{subsec:ntile_out}
For the potential exterior to the tile, but near it, we now want to compute
\be \chi_{\rm N}(x,y,z)
= \frac{\bar B_z}{2 \pi}
\int_{-\Delta s/2}^{\Delta s/2} {\rm d}x' \int_{-\Delta s/2}^{\Delta s/2} {\rm d}y'
\frac{1}{\sqrt{(x - x')^2 + (y - y')^2 + z^2}}
~, \ee
where, as above, the arguments of $\chi_{\rm N}$ refers to distance with respect to the center of the source pixel, which we take as the origin. We change variables {\it via} Equations (\ref{eqn:xpm}) and (\ref{eqn:ypm}), so the upper and lower limits of integration $\pm \Delta/2$ for $x'$ become $x_\mp$ for $x''$, and similarly for $y'$ and $y''$, so
\be \chi_{\rm N}(x,y)
= \frac{\bar B_z}{2 \pi} \int_{x_-}^{x_+} {\rm d}x'' \int_{y_-}^{y_+} {\rm d}y''
\frac{1}{\sqrt{x''^2 + y''^2 + z^2}} ~. \ee
Defining $a = (x''^2 + z^2)$,
\bea \chi_{\rm N}(x,y)
&=& \frac{\bar B_z}{2 \pi} \int_{x_-}^{x_+} {\rm d}x'' \int_{y_-}^{y_+} {\rm d}y''
\frac{1}{\sqrt{y''^2 + a^2}} \\
&=& \frac{\bar B_z}{2 \pi} \int_{x_-}^{x_+} {\rm d}x'' \left .
\ln [\sqrt{y''^2 + a^2} + y''] \right \vert_{y_-}^{y_+} \\
&=& \frac{\bar B_z}{2 \pi} \int_{x_-}^{x_+} {\rm d}x'' \left (
\ln [\sqrt{x''^2 + (y + \Delta s/2)^2 + z^2} + (y + \Delta s/2)]
\right . \nn \\
&&
\left .
- \ln [\sqrt{x''^2 + (y - \Delta s/2)^2 + z^2} + (y - \Delta s/2)] \right )
~. \label{eqn:neum_pot_halted} \eea
We had no success at further reducing this expression
analytically, \footnote{By differentiating Equation
  (\ref{eqn:neum_greens}) prior to integrating the source distribution
  for a tile, Barnes {\it et al.} (2009) were able to derive an
  analytic expressions for the field's components; see
  http://www.cora.nwra.com/$\sim$graham/Potential/ } so instead
evaluated it numerically for points in the neighborhood of the
tile. For five classes of nearby, off-tile points, discrepancies
between the point-source potential, Equation (\ref{eqn:neum_greens}),
and our numerical results were above 1.5\%. For these sets of points,
we used our results to derive coefficients to ``correct'' the distance
factor in the denominator of Equation (\ref{eqn:neum_greens}). For two
points directly above the tile's center, at $z=1$ and $z=2$, the
denominator should be multiplied by 0.9286 and 0.9800,
respectively. For the four nearest neighbors in $x$ and $y$, the
denominator should be multiplied by 1.0381 for $z=0$ and 0.9878 for
$z=1$. Finally, for the four corner neighbors in $x$ and $y$ at $z=0$,
the denominator should be multiplied by 1.0249.

Note that using these ``corrections'' for $0 < z \le 2$ will produce a Neumann potential that disagrees in these zones with the Dirichlet potential, computed {\it via} Equation (\ref{eqn:dirichlet_3d}), extrapolated from the Neumann potential at $z=0$.

\vfill

\eject

\bibliographystyle{spr-mp-sola}

%\end{article}
\end{document}